%% file: main.tex
\PassOptionsToPackage{table}{xcolor}
\documentclass[sigconf]{acmart}
\AtBeginDocument{%
  }

\copyrightyear{2026}
\acmYear{2026}
\setcopyright{cc}
\setcctype{by}
\acmConference[ICSE '26]{2026 IEEE/ACM 48th International Conference on Software Engineering}{April 12--18, 2026}{Rio de Janeiro, Brazil}
\acmBooktitle{2026 IEEE/ACM 48th International Conference on Software Engineering (ICSE '26), April 12--18, 2026, Rio de Janeiro, Brazil}
\acmPrice{}
\acmDOI{10.1145/3744916.3773140}
\acmISBN{979-8-4007-2025-3/26/04}

\usepackage{array}
\newcolumntype{C}[1]{>{\centering\arraybackslash}p{#1}}

\usepackage[shortlabels]{enumitem}

\definecolor{mattegreen}{RGB}{146,174,110} 
\newcolumntype{o}{>{\columncolor{mattegreen!20}}c} 
\definecolor{deeporange}{RGB}{180,90,0} 
\definecolor{darkgreen}{RGB}{0,160,0}

\usepackage{listings}

\usepackage{textcomp} 

\newcommand{\mycircle}[1]{\textcircled{\footnotesize #1}}
\newcommand{\mycircleT}[1]{\textcircled{\footnotesize #1}}
\newcommand{\mycircleS}[1]{\textcircled{\footnotesize #1}}

\colorlet{shadecolor}{gray!10}

\newenvironment{answerbox}[1][\linewidth]{%
  \begin{shaded}
  \setlength{\parindent}{0pt}
}{%
  \end{shaded}
}

\usepackage{framed}
\newenvironment{answer_env}[2]{%
  \begin{answerbox}[\linewidth]
  \textbf{Answer for RQ#1:}\ #2
}{%
  \end{answerbox}
}

\makeatletter
\renewcommand{\section}{%
  \@startsection{section}{1}{0pt}%
    {-4pt plus -1pt minus -1pt}
    {2pt plus 1pt minus 1pt}
    {\normalfont\large\bfseries}}    
    
\renewcommand{\subsection}{%
  \@startsection{subsection}{2}{0pt}%
    {-4pt plus -1pt minus -1pt}
    {2pt plus 1pt minus 1pt}
    {\normalfont\normalsize\bfseries}}

\renewcommand{\subsubsection}{%
  \@startsection{subsubsection}{3}{0pt}%
    {4pt plus 1pt minus 1pt}%
    {-0.5em}
    {\normalfont\normalsize\bfseries}}
\makeatother

\begin{document}
\title[A Process-Oriented Error Analysis of Software Development Agents in Real-World GitHub Scenarios]{Beyond Final Code: A Process-Oriented Error Analysis of Software Development Agents in Real-World GitHub Scenarios}

\author{Zhi Chen}
\affiliation{%
 \institution{The Centre for Research on Intelligent Software Engineering (RISE), Singapore Management University}
  \country{Singapore}
}
\email{zhi.chen.2023@smu.edu.sg}

\author{Wei Ma}
\authornote{Corresponding author.}
\affiliation{%
 \institution{The Centre for Research on Intelligent Software Engineering (RISE), Singapore Management University}
  \country{Singapore}
}
\email{weima@smu.edu.sg}

\author{Lingxiao Jiang}
\affiliation{%
  \institution{The Centre for Research on Intelligent Software Engineering (RISE), Singapore Management University}
  \country{Singapore}
}
\email{lxjiang@smu.edu.sg}

\begin{abstract}
\input{sections/abstract}
\end{abstract}



\maketitle
\input{sections/introduction}

\input{sections/study_design}
\input{sections/research_questions}
\input{sections/exploratory_analysis}

\input{sections/solving_phase_error_prevalence_analysis}
\input{sections/solving_phase_challenging_error_analysis}

\input{sections/testing_phase_error_analysis}
\input{sections/discussion}

\input{sections/related_work}

\input{sections/conclusion}

\begin{acks}
This research is supported by the Ministry of Education, Singapore under its Academic Research Fund Tier 3 (Award ID: MOET32020-0004). Any opinions, findings and conclusions or recommendations expressed in this material are those of the author(s) and do not reflect the views of the Ministry of Education, Singapore.
\end{acks}

\balance
\bibliographystyle{ACM-Reference-Format}
\bibliography{main}

\end{document}

%% file: sections/abstract.tex
AI-driven software development has rapidly advanced with the emergence of software development agents that leverage large language models (LLMs) to tackle complex, repository-level software engineering tasks. These agents go beyond just generation of final code; they engage in multi-step reasoning, utilize various tools for code modification and debugging, and interact with execution environments to diagnose and iteratively resolve issues. However, most existing evaluations focus primarily on static analyses of final code outputs, yielding limited insights into the agents' dynamic problem-solving processes. To fill this gap, we conduct an in-depth empirical study on 3,977 solving-phase trajectories and 3,931 testing-phase logs from 8 top-ranked agents evaluated on 500 GitHub issues in the SWE-Bench benchmark. 
Our exploratory analysis shows that Python execution errors during the issue resolution phase correlate with lower resolution rates and increased reasoning overheads. We have identified the most prevalent errors---such as \textit{ModuleNotFoundError} and \textit{TypeError}---and highlighted particularly challenging errors like \textit{OSError} and database-related issues (e.g., \textit{IntegrityError}) that demand significantly more debugging effort. Furthermore, we have discovered 3 bugs in the SWE-Bench platform that affect benchmark fairness and accuracy; these issues have been reported to and confirmed by the maintainers. To promote transparency and foster future research, we publicly share our datasets and analysis scripts.

%% file: sections/introduction.tex
\section{Introduction}
\label{sec:intro}

The field of AI-based automatic software engineering is undergoing a transformative shift with the emergence of software development agents~\cite{he2024llm,chen2024evaluating}. Building on this shift, recent years have seen rapid evolution in AI techniques for software engineering. Early approaches relied on traditional machine learning and deep learning models to perform tasks such as bug detection~\cite{kharkar2022learning,li2020improving}, vulnerability classification~\cite{marjanov2022machine}, and automatic program repair~\cite{li2020dlfix,jiang2021cure,chen2019sequencer}. The advent of generative code models marked a significant turning point: early models like CodeBERT~\cite{feng-etal-2020-codebert} and CodeT5~\cite{wang2021codet5}, fine-tuned on source code, laid the groundwork for more advanced systems. Today, large language models (LLMs) with billions of parameters—such as DeepSeek R1~\cite{guo2025deepseek}, Meta Llama 3~\cite{grattafiori2024llama}, OpenAI ChatGPT~\cite{achiam2023gpt}, and Google Gemini~\cite{team2023gemini}—have demonstrated extraordinary capabilities. These models excel at tasks including code generation~\cite{shinn2023reflexion}, program repair~\cite{xia2024automated}, code summarization~\cite{ahmed2022few}, and automated code review~\cite{lu2023llama}, as evidenced by benchmarks such as EvalPlus~\cite{liu2023your} and BigCodeBench~\cite{zhuo2025bigcodebench}. However, simple function generation tasks are increasingly viewed as inadequate for capturing the complexities of real-world software engineering~\cite{jimenez2024swebench}. In response, researchers have equipped LLMs with agentic workflows that enable interaction with external environments and tools, thereby enhancing their ability to tackle more complex tasks~\cite{jin2024llms}. Early software development agents—such as Devin\footnote{\url{https://www.cognition.ai/blog/introducing-devin}}—gained considerable attention at their debut. Since then, a growing number of agents from both industry~\cite{ma2024understand,wang2024openhands,liu2024marscode,ma2024lingma} and academia~\cite{yang2024swe,hongjinlearn,xia2024agentless} have emerged, demonstrating promising results by autonomously addressing repository-level challenges.

\begin{figure*}[t]
    \centering
    \scalebox{1.0}{
    \includegraphics[width=\textwidth]{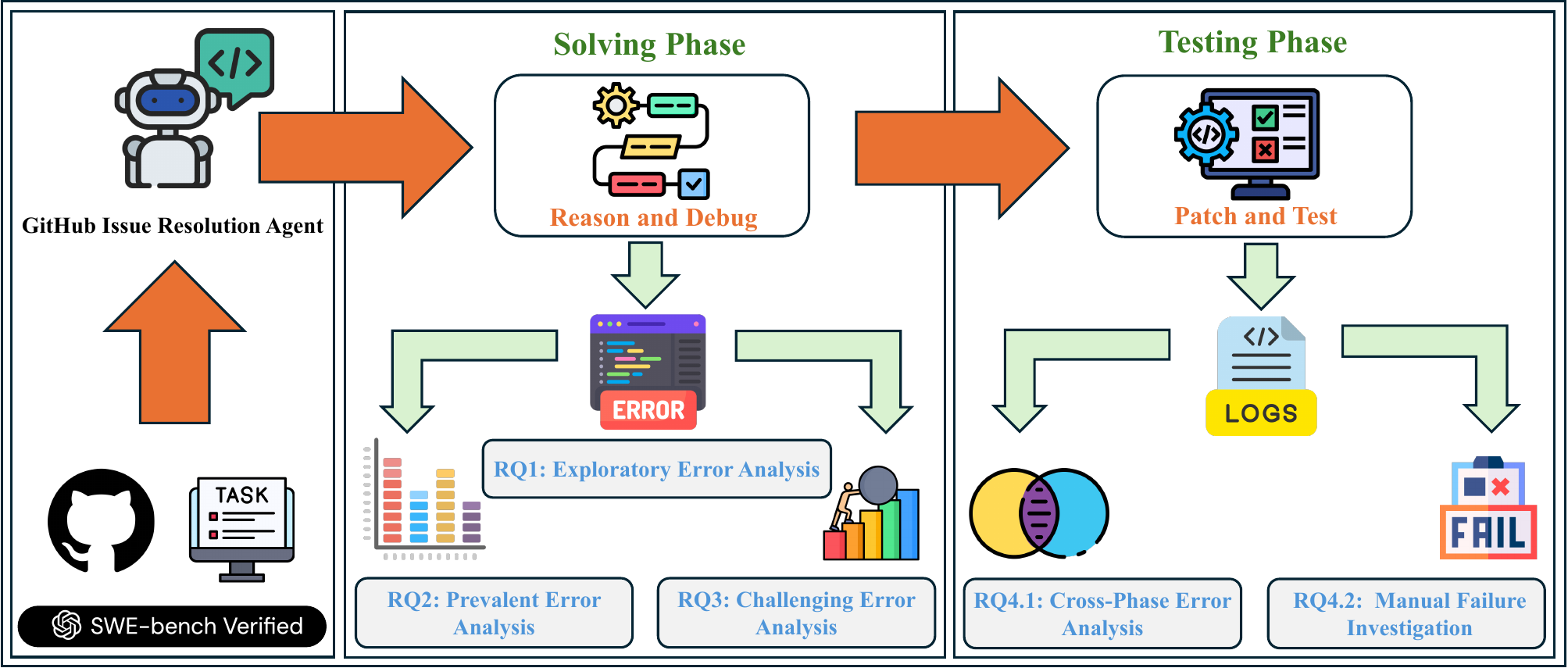}}
      \caption{Study overview: solving-phase trajectories inform analyses of unexpected-error impact (RQ1), common-error prevalence (RQ2), and challenging-error identification (RQ3); testing-phase logs reveal testing errors and failures(RQ4).}
    \label{fig:overview}
\end{figure*}

However, current evaluations of AI-based software development primarily focus on their final outputs. Several studies~\cite{liu2024no,liu2024refining,rabbi2024ai,siddiq2024quality,chen2024evaluating,pearce2022asleep,majdinasab2024assessing,asare2023github,hamer2024just,nguyen2022empirical} have assessed AI-driven software development by examining the final code produced by LLMs and software development agents, relying on static metrics such as bug counts for reliability~\cite{zhong2024can}, vulnerability assessments for security~\cite{liu2024no}, and code duplication or complexity measures for maintainability~\cite{chen2024evaluating}. While these evaluations provide valuable insights, they offer only a partial view of an agent’s true capabilities and limitations,
as in the real-world software development, agents engage in iterative, dynamic processes to solve tasks, often involving multiple rounds of experimentation, debugging, and adaptation~\cite{yang2024swe,chen2024evaluating,he2024llm}. To the best of our knowledge, no work has yet studied the errors occurring during the resolution processes in the context of agents.
This gap underscores the importance of analyzing process-oriented data to better assess agents' capabilities and limitations.


To fill this gap, we analyze agents’ resolution process data from SWE-Bench~\cite{yang2024swe}, a widely adopted benchmark for assessing AI software developers' ability to automatically resolve GitHub issues\footnote{https://openai.com/index/introducing-swe-bench-verified/}. Figure~\ref{fig:overview} presents an overview of our study, which leverages 3,977 issue-solving-phase trajectory files and 3,931 testing-phase logs collected from 8 top-ranked agents addressing 500 real-world GitHub issues across 12 popular Python repositories. We first conduct an exploratory analysis to investigate how Python execution errors during the solving phase influence both final patch quality and agent problem-solving steps. Building on these insights, we then examine the overall prevalence of error types in real-world GitHub tasks and identify which of these errors are particularly challenging for agents to resolve. In addition, we analyze the testing-phase logs to uncover the specific reasons behind patch failures and to determine how many errors observed during the resolution phase persist into final patch failures. Overall, our study offers a new perspective on agent performance—from initial code modifications through final validation—and provides guidance for future improvements in error-handling capabilities.

In summary, our work makes four main contributions:
\begin{enumerate}[1),leftmargin=1em]
    \item This study is the first to jointly analyze issue-solving-phase trajectories and testing logs for Github resolution agents, moving beyond prior work that focused only on final code solutions.
    
    \item We identify the most prevalent and challenging errors encountered by these agents, highlighting critical areas where improvements in error handling and recovery are urgently needed.

    \item Our analysis of unresolved task failures reveals 3 critical bugs in the SWE-Bench platform that compromise the benchmark’s correctness and fairness. All 3 bugs were reported to and confirmed by the SWE-Bench authors.
    
    \item To promote transparency and reproducibility, we publicly share our datasets and scripts\footnote{\url{https://figshare.com/s/bf7c3e656d1a57e1f50b}} to support further research.
\end{enumerate}

The rest of this paper is organized as follows. Section~\ref{sec:study_design} presents the SWE benchmark, the code agents used, and our data collection process. Section~\ref{sec:RQs} introduces the four research questions (RQs) we investigate: the impact of unexpected errors (RQ1), the prevalence of common errors (RQ2), the identification of challenging errors (RQ3), and an analysis of patch failures and testing errors(RQ4). Sections~\ref{sec:exploratory_analysis}, \ref{sec:error_prevalence_analysis}, \ref{sec:challenging_errors_analysis}, and \ref{sec:patch_failure} provide the analysis results and answers to these questions, respectively. Section~\ref{sec:discussion} discusses the implications of our findings, outlines directions for future research, and examines potential threats to validity, while Section~\ref{sec:related_work} reviews related work. Finally, Section~\ref{sec:conclusion} concludes the study.

%% file: sections/study_design.tex
\section{Study Design}
\label{sec:study_design}

\subsection{Choice of SWE-Bench Verified}
We selected SWE-Bench Verified~\cite{yang2024swe} as the source of our study data to evaluate AI software development agents on real-world GitHub issues. The benchmark is built on 500 pairs of Issues and Pull-Requests from 12 validated open-source Python repositories;
these issue resolution tasks were verified by professional software engineers in collaboration with the OpenAI preparedness team. Unlike simpler code generation benchmarks (e.g., HumanEval~\cite{chen2021evaluating}, MBPP~\cite{austin2021program}, or BigCodeBench~\cite{zhuo2025bigcodebench}), SWE-Bench Verified presents realistic repository-level challenges.

For each task in SWE-Bench Verified, an agent receives an issue description for a project and the corresponding codebase of the project, then engages in a multi-step reasoning and resolution process that includes tool-assisted code modifications, debugging, and interactions with execution environments, in order to produce code patches for the task~\cite{yang2024swe}.
Detailed trajectory files capture their internal reasoning and tool usages, while final logs document the patch correctness
against test cases. This comprehensive process data enables our in-depth analysis of agent performance and their underlying error patterns, directly addressing the evaluation gap discussed in our introduction.

\subsection{Choice of Agents}
\label{sec:choice_agents}

We first obtained 24 unique agents by filtering duplicate submissions (retaining only the best version per unique agent) from the top 30 agents on the leaderboard by resolution rate.\footnote{\url{https://www.swebench.com/\#verified}} We then applied the following criteria to identify agents whose trajectories are best suited for error mining:

\noindent\textbf{1. Inclusion of Execution Observation.} Trajectories must record execution-related outputs (e.g., running \textit{python reproduce.py}) with sufficient details such as error messages, state transitions, or debugging outputs. For example, Learn-by Interact\cite{su2025learn} omits bash-tool details—neither executed commands nor stdout/stderr—making analysis infeasible. 

\noindent\textbf{2. Clear Observation Delineation.} Observation sections must have clear boundaries (e.g., marked by \texttt{OBSERVATION} or \texttt{RESPONSE}) to reliably identify relevant information. 

\noindent\textbf{3. Unmodified Contents.} Observations should retain their original outputs produced by tools or environment used by the agents, preserving wording and formatting for accurate error extraction.

Following these criteria, we selected 8 diverse agents for our study. Table~\ref{tab:selected_agents} summarizes them with key metadata\footnote{From each agent’s metadata.yaml file.}.

\begin{table}[h]
\centering
\caption{Selected Agents}
\label{tab:selected_agents}
\resizebox{\columnwidth}{!}{%
\begin{tabular}{@{}lllccc@{}}
\toprule
No. & Rank & Model                        & \% Resolved & Date       & Base LLM                 \\ \midrule
1   & 1    & W\&B Programmer             & 64.60       & 2025-01-17 & o1-2024-12-17            \\
2   & 2    & Blackbox AI Agent           & 62.80       & 2025-01-10 & Unknown                  \\
3   & 5    & Devlo                       & 58.20       & 2024-12-13 & Unknown        \\
4   & 12   & CodeAct                     & 53.00       & 2024-10-29 & Claude 3.5 Sonnet        \\
5   & 14   & Engine Labs                 & 51.80       & 2024-11-25 & Claude 3.5 Sonnet        \\
6   & 17   & Bytedance MarsCode Agent    & 50.00       & 2024-11-25 & Unknown                  \\
7   & 19   & Tools                       & 49.00       & 2024-10-22 & Claude 3.5 Sonnet        \\
8   & 26   & CodeShell              & 44.20       & 2025-01-18 & Gemini-2.0-flash-exp     \\
\bottomrule
\end{tabular}%
}
\footnotesize
\begin{flushleft}
\textbf{Note:} The data used in this study is up-to-date as of January 20, 2025.
\end{flushleft}
\end{table}

\paragraph{\textbf{Impact of Backbone Models}}
Prior studies~\cite{yang2024unveiling,chen2024promise,kong2025demystifying} show that code LMs can memorise—and potentially leak—training snippets.  
If such leakage were the dominant factor, agents sharing the same backbone would perform similarly.  
Yet on SWE-Bench three Claude-3.5 Sonnet agents score very differently: \textit{CodeStory} 62.2\%, \textit{AutoCodeRover-v2.0}~\cite{ruan2024specrover} 46.2\%, and \textit{SWE-Agent}~\cite{yang2024swe} 33.6\%. This gap indicates that results and trajectories are shaped by the \emph{entire agentic stack}—backbone LM, workflow design, and tool set—rather than by the model alone.  

\subsection{Data Collection} 
\label{subsec:data_collection}

\paragraph{\textbf{Trajectories of Issue-Solving Phases}} 
To understand how software development agents tackle SWE-Bench tasks, each patch submission from an agent includes a trajectory file (e.g., \lstinline{astropy__astropy-8797.json}) that records the agent’s intermediate steps and decisions during task resolution. This file provides insight into the agent's internal workflow and the emergence of errors. Although these files can be formatted in various ways (e.g., JSON, YAML, or Markdown), they are designed to capture the task progression step by step—from issue reproduction and code modification through test execution and the associated execution observations, culminating in the final patch submission. We collected 3,977 trajectories from 8 selected agents across 500 tasks. 


\begin{table}[h]
  \caption{File Counts for Each Agent}
  \label{tab:agents_files}
  \centering
  \resizebox{0.8\columnwidth}{!}{%
    \begin{tabular}{lcc}
      \toprule
      Agent           & Solving Phase Trajectories & Testing Phase Logs \\
      \midrule
      W\&B Programmer & 500   & 499  \\
      Blackbox AI     & 480   & 480  \\
      Devlo           & 500   & 500  \\
      CodeAct         & 500   & 493  \\
      Engine Labs     & 499   & 492  \\
      MarsCode        & 500   & 498  \\
      Tools           & 498   & 483  \\
      CodeShell       & 500   & 486  \\
      \midrule
      \textbf{Total}  & \textbf{3,977}  & \textbf{3,931} \\
      \bottomrule
    \end{tabular}%
  }
\end{table}


\paragraph{\textbf{Logs of Testing Phases}} 
After an agent submits a patch, SWE-Bench applies the patch to the project repository and runs a suite of test cases to determine whether the patch resolves the issue. Test logs are typically stored in \texttt{logs/output.txt} and generated via the \texttt{eval.sh} script on the \texttt{patch.diff} file; the logs provide quantitative measures of patch effectiveness and capture errors during testing. We collected all available logs from the eight selected agents across 500 tasks, resulting in a total of 3,931 logs as shown in Table \ref{tab:agents_files}.


\textit{\textbf{Error Data Extraction.}} Due to the absence of a common trajectory format, we use agent-specific parsers to identify tool execution output using markers like \texttt{OBSERVATION}, and extract both Python error types and messages via regular expressions that match tokens ending in “Error:” and capture the subsequent message text.

%% file: sections/research_questions.tex
\section{Research Questions}
\label{sec:RQs}

Figure~\ref{fig:overview} provides an overview of our error analysis study. The agents' workflows for addressing issues are divided into two stages, the solving phase and the testing phase. Our study follows this workflow and analyzes errors at each stage using a range of techniques, from basic statistics to in-depth examinations guided by four research questions (RQs). Specifically, we begin with a general exploratory error analysis (RQ1), then identify the prevalent and challenging errors (RQ2 and RQ3) to better understand the execution errors that arise during the agents' iterative reasoning and debugging process. Finally, we perform a cross-phase error analysis (RQ4.1) and manually investigate patch failures (RQ4.2).

\subsection{RQ1: Exploratory Error Analysis}
\textbf{Motivation:}  
SWE-Bench tasks, derived from real-world GitHub issues, require agents to understand the problem description and potentially many code files in the project, and identify and modify relevant code files, making them significantly more complex than isolated function-level code generation or bug fixing. A typical workflow for resolving such issues involves issue reproduction, iterative code modifications, and running tests to validate patches. 
During this multi-step process for issue solving, Python execution failures manifested as parsing or runtime errors would disrupt planned modifications, introduce additional repair tasks, and ultimately degrade the quality of generated patches.
Our exploratory analysis examines how the frequency and nature of such failures impact agents' reasoning trajectories and their final patch quality.

\textbf{RQ1:} \textit{How do Python execution failures during the issue-solving phase affect the final patch quality and reasoning steps of software development agents?}

\subsection{RQ2: Prevalent Error Analysis}

\textbf{Motivation:}  
Building on RQ1, 
we examine the prevalence of different error types encountered by agents when solving real-world GitHub issues.
Given the complexity of these tasks, which require understanding entire project-level code repositories, iterative modifications, and extensive debugging, it is crucial to identify the errors that occur most frequently during the debugging and reasoning processes, such as those encountered during Python code execution for verifying the correctness of modified code files.
By identifying these prevalent errors, we can reveal the current limitations of state-of-the-art agents and provide targeted guidance to mitigate these pitfalls, thereby improving issue resolution success rate and efficiency while reducing computational resource demands.

\textbf{RQ2:} \textit{Which error types are most commonly encountered by software development agents during the debugging and reasoning process when solving real-world GitHub issues?}

\subsection{RQ3: Challenging Error Analysis}

\textbf{Motivation:}  
While RQ2 quantified the prevalence of various Python execution errors encountered by software development agents, it remains unclear which error types significantly disrupt the agents' reasoning and debugging processes, that is, which errors are particularly challenging to recover from. Although many errors occur frequently, some can be resolved with simple fixes, whereas others require more sophisticated intervention and may be difficult to correct in a single attempt. Identifying these challenging errors is essential for understanding current limitations in agent-based debugging and for guiding the development of improved diagnostic and recovery strategies.

\textbf{RQ3:} \textit{Which error types are particularly challenging for software development agents to recover from during the GitHub Issue solving process?}

\subsection{RQ4: Failure and Cross-Phase Error Analysis}

\textbf{Motivation:}
While our previous research questions have examined errors during the GitHub issue-solving phase, many failures occur after agents submit their patches. After completing the solving phase—or when they believe they have completed it—agents submit their generated patches, which are then applied to the project and evaluated by SWE-Bench using test cases, yielding a binary “resolved” or “not resolved” outcome. However, this binary evaluation does not reveal the underlying causes of these failures. Thus, it is essential to delve deeper into testing-phase errors and explore whether any defect in the SWE-Bench evaluation platform might also contribute to failures. Furthermore, patches may fail due to Python execution errors observed during the solving phase that were not addressed, which we refer to as \textit{cross-phase errors}. Identifying these failures and understanding their underlying causes is crucial for uncovering the limitations of current agents' error recovery mechanisms and emphasizes the need for enhanced error analysis in developing more robust automated software agents.

\textbf{RQ4:} \textit{What are the underlying failure reasons for patches, and which errors encountered during the testing phase were previously observed (but not resolved) during the solving phase?}

%% file: sections/exploratory_analysis.tex
\section{Exploratory Error Analysis}
\label{sec:exploratory_analysis}

\subsection{Experimental Setup}
To answer 
``\textbf{RQ1}: \textit{How do Python execution failures during the task-solving phase affect the final patch quality and reasoning steps of software development agents?}'' 
we conduct an exploratory analysis from three perspectives:

\noindent\textbf{1. Impact of Error Occurrence on Resolution Rate.}  
We compare task instances (a task instance represents an issue resolution by an agent) where agents encountered any errors against those with no errors. This metric isolates whether the mere presence of execution failures influences the final patch resolution rate (the proportion of task instances whose submitted patches successfully pass all associated tests)
, providing a baseline understanding of their effect.

\noindent\textbf{2. Impact of Error Frequency on Resolution Rate.}  
We group task instances based on the number of errors occurred during resolution and compare the corresponding resolution rates. By doing so, we analyze the cumulative effect of errors, checking whether tasks with more execution failures correlate with lower resolution rates.

\noindent\textbf{3. Correlation with Reasoning Steps.}  
We use Pearson correlation analysis~\cite{cohen2009pearson} to examine the relationship between the number of errors and the reasoning steps per task. 
We expect that tasks with more execution failures will require additional steps for error diagnosis and repair. At the same time, an excessive number of reasoning steps can enlarge the overall context, potentially degrading performance and leading to further errors. This analysis is only a statistic descriptive probe: it quantifies the correlation between error frequencies and reasoning-step counts, indicating whether longer trajectories correlate with more errors. It is \emph{not} a predictive or causal analysis.

\subsection{Experimental Results}

\paragraph{\textbf{Impact of Error Occurrence}} Table~\ref{tab:resolution_rate} shows the resolution rates for tasks where agents encountered errors during resolution versus those without errors. Overall, the resolution rates are close, 54.61\% for tasks with errors compared to 54.42\% for tasks without errors, indicating that merely encountering an error during the solving process does not significantly affect final patch quality. While some agents show slightly larger differences (e.g., Blackbox AI's 61.49\% for tasks with errors and 72.51\% for tasks without errors), the overall trend suggests that agents generally are capable of learning from error feedback to resolve issues and complete tasks. However, the extent to which agents can tolerate errors remains unclear. Our next step examines how error frequency influences resolution rates, which may reveal more nuanced effects on agent performance.

\begin{table}[ht]
\caption{Resolution Rate by Agent and Error Occurrence}
\label{tab:resolution_rate}
\centering
\resizebox{0.8\columnwidth}{!}{%
\begin{tabular}{lcc}
\toprule
\textbf{Agent} & \textbf{With Error} & \textbf{Without Error} \\
\midrule
W\&B Programmer & 63.55\% (68/107)   & 64.89\% (255/393) \\
Blackbox AI     & 61.49\% (190/309)  & 72.51\% (124/171) \\
Devlo           & 56.85\% (191/336)  & 60.98\% (100/164) \\
CodeAct         & 52.72\% (155/294)  & 53.40\% (110/206) \\
Engine Labs     & 54.33\% (163/300)  & 48.24\% (96/199)  \\
MarsCode        & 55.81\% (120/215)  & 45.61\% (130/285) \\
Tools           & 49.63\% (133/268)  & 48.70\% (112/230) \\
CodeShell       & 41.03\% (64/156)   & 45.64\% (157/344) \\
\midrule
\textbf{Overall} & 54.61\% (1084/1985) & 54.42\% (1084/1992) \\
\bottomrule
\end{tabular}
}
\end{table}

\begin{figure}[ht]
\centering
\includegraphics[width=\columnwidth]{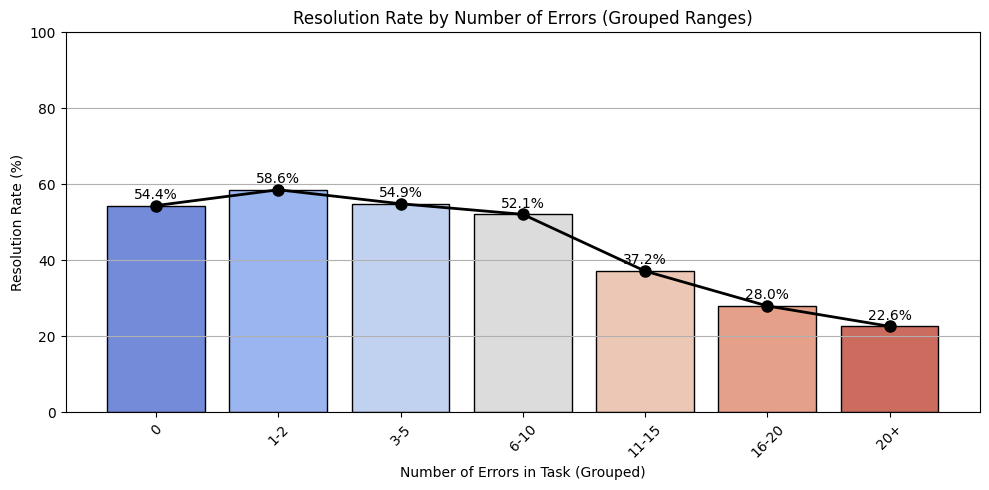} 
\caption{Resolution Rate by Error Frequency}
\label{fig:error_frequency}
\end{figure}

\paragraph{\textbf{Impact of Error Frequency}}Figure~\ref{fig:error_frequency} shows that task instances with only a couple of errors (1--2 errors) during resolution actually yield a slightly higher resolution rate (58.6\%) compared to task instances with no errors (54.4\%). However, as the number of errors increases, the resolution rate declines gradually, dropping to 54.9\% for 3--5 errors and 52.1\% for 6--10 errors. When the number of errors reaches 11--15, the resolution rate dramatically falls to 37.2\%, and further declines to 28.0\% (16--20 errors) and 22.6\% (20+ errors). This suggests that while agents are capable of resolving a small number of errors, an excessive number of errors significantly harms their ability in producing correct patches.
This indicates that current agents struggle with error handling, still far from the ideal scenario where they can effectively learn from diverse error messages and successfully complete their tasks.

\paragraph{\textbf{Correlation with Reasoning Steps}} 
Table~\ref{tab:correlation} presents the Pearson correlation coefficients between the number of errors encountered and the number of reasoning steps taken by the agents.
The analysis reveals a significant positive relationship (Overall Pearson $r = 0.59423$, $p < 0.001$), suggesting that as agents encounter more errors, they tend to take additional reasoning steps to recover or address these errors; however, this extra reasoning may increase the complexity of the process and potentially lead to further errors and a negative feedback loop.
These findings underscore the importance of reducing error frequency through improved error recovery strategies, which could streamline the reasoning process, enhance patch success rates, and contribute to more efficient and sustainable software development practices. 

\begin{table}[h]
\centering
\caption{Correlation between Errors and Steps}
\label{tab:correlation}
\resizebox{0.8\columnwidth}{!}{%
\begin{tabular}{lcc}
\toprule
\textbf{Agent}      & \textbf{Pearson Corr.} & \textbf{p-value} \\
\midrule
W\&B Programmer     & 0.69898              & $5.66\times10^{-17}$ \\
Blackbox AI         & 0.53433              & $3.28\times10^{-24}$ \\
Devlo               & 0.40917              & $5.40\times10^{-15}$ \\
CodeAct             & 0.55893              & $1.48\times10^{-25}$ \\
Engine Labs         & 0.37422              & $2.09\times10^{-11}$ \\
MarsCode            & 0.94036              & $1.04\times10^{-101}$ \\
Tools               & 0.79235              & $5.00\times10^{-59}$ \\
CodeShell           & 0.48887              & $9.48\times10^{-11}$ \\
\midrule
\textbf{Overall}    & \textbf{0.59423}     & \textbf{$8.21\times10^{-190}$} \\
\bottomrule
\end{tabular}%
}
\footnotesize
\begin{flushleft}
\textbf{Note:} The p-value indicates the probability that the observed correlation occurred by chance. A p-value below 0.05 is considered statistically significant.
\end{flushleft}
\end{table}
%


Our investigation reveals three key insights: (1) The mere occurrence of an error during the resolution does not show significant compromise on patch quality. (2) Increasing number of errors is associated with a marked decline in patch quality. (3) Higher error frequency correlates with increased corrective reasoning, complicating patching process and raising computational costs.

\begin{answer_env}{1}{Increased error frequency is associated with lower-quality patches, as reflected by a lower resolution rate, and with more reasoning steps, which in turn complicate patching and raise computational costs.}
\end{answer_env}

%% file: sections/solving_phase_error_prevalence_analysis.tex
\section{Prevalent Error Analysis}
\label{sec:error_prevalence_analysis}

\subsection{Experimental Setup}
To address 
``\textbf{RQ2:} \textit{Which error types are most commonly encountered by software development agents during the debugging and reasoning process when solving real-world GitHub issues?}'' 
we analyze the prevalence of various Python execution errors occurred during resolution using our trajectory dataset described in Section~\ref{subsec:data_collection}. Our analysis quantifies error prevalence using three metrics:
\begin{itemize}[leftmargin=*]
    \item \textbf{Total Occurrence Count.} The raw frequency of each error type across all tasks and agents.
    \item \textbf{Task-Level Prevalence.} The number of tasks ("task" refer to a github issue) in which a specific error type occurs at least once.
    \item \textbf{Agent-Level Prevalence.} The number of agents that exhibit a specific error type in at least one of their tasks.
\end{itemize}

We extracted error lines from the \textit{OBSERVATION} sections (see Section~\ref{sec:choice_agents}), which record the execution outputs of Python code in the solving-phase trajectory files. We then applied a combination of regular expressions and keyword matching to classify each error into specific types (e.g., \texttt{SyntaxError}, \texttt{ModuleNotFoundError}). Finally, we aggregated the data using the three metrics, providing 
insights into
the prevalence of common errors encountered during GitHub issue resolution.

\subsection{Experimental Results}
Our initial analysis identified a total of 92 non-generic error types. To focus on the prevalent errors and avoid those specific to only a few tasks or agents, we set a threshold of at least 5 occurrences at the task level (Task Prev) and at least 4 occurrences at the agent level (Agent Prev). After applying these filters, we retained 32 error types, as shown in Table ~\ref{tab:prevalent-errors}. Our analysis of these 32 error types reveals two overarching challenges faced by the agents. We group the errors into two main categories—\textbf{\textit{Python Built-in Errors}} and \textbf{\textit{Custom-Defined Exceptions}}—each further subdivided to highlight the specific nature of the issues.

\begin{table}[h]
  \caption{Prevalent Errors}
  \label{tab:prevalent-errors}
  \centering
  \resizebox{\columnwidth}{!}{%
    \begin{tabular}{r l o r r}
      \toprule
      No. & Error Type & \textbf{\textcolor{deeporange}{Total Occ. $\downarrow$}}  & Task Prev. & Agent Prev. \\
      \midrule
       \mycircle{1} & ModuleNotFoundError                             & 1053 & 267 & 8 \\
       \mycircle{2} & TypeError                                       &  992 & 217 & 8 \\
       \mycircle{3} & AttributeError                                  &  841 & 233 & 8 \\
       \mycircle{4} & django.db.utils.OperationalError                &  528 &  99 & 8 \\
       \mycircle{5} & sqlite3.OperationalError                        &  503 &  94 & 8 \\
       \mycircle{6} & ImportError                                     &  462 & 183 & 8 \\
       \mycircle{7} & ValueError                                      &  436 & 113 & 8 \\
       \mycircle{8} & NameError                                       &  256 & 105 & 8 \\
       \mycircle{9} & KeyError                                        &  209 &  63 & 8 \\
      \mycircleS{10} & django.db.utils.IntegrityError                  &  168 &  14 & 6 \\
      \mycircleS{11} & UnicodeDecodeError                              &  166 &  43 & 6 \\
      \mycircleS{12} & SyntaxError                                     &  163 &  67 & 8 \\
      \mycircleS{13} & sqlite3.IntegrityError                          &  119 &  14 & 5 \\
      \mycircleS{14} & django.core.exceptions.FieldError               &  107 &  24 & 7 \\
      \mycircleS{15} & IndexError                                      &   93 &  25 & 7 \\
      \mycircleS{16} & FileNotFoundError                               &   88 &  59 & 7 \\
      \mycircleS{17} & django.core.exceptions.ValidationError          &   79 &  10 & 5 \\
      \mycircleS{18} & UnicodeEncodeError                              &   40 &  19 & 5 \\
      \mycircleS{19} & IndentationError                                &   36 &  20 & 6 \\
      \mycircleS{20} & OSError                                         &   32 &   6 & 5 \\
      \mycircleS{21} & UnboundLocalError                               &   32 &  11 & 7 \\
      \mycircleS{22} & django.core.management.base.SystemCheckError    &   30 &  22 & 7 \\
      \mycircleS{23} & RecursionError                                  &   29 &  11 & 5 \\
      \mycircleS{24} & django.core.management.base.CommandError        &   29 &  14 & 5 \\
      \mycircleS{25} & subprocess.CalledProcessError                   &   29 &   7 & 7 \\
      \mycircleS{26} & LookupError                                     &   25 &  14 & 8 \\
      \mycircleS{27} & django.db.utils.ProgrammingError                &   22 &   6 & 4 \\
      \mycircleS{28} & django.db.utils.NotSupportedError               &   20 &  11 & 6 \\
      \mycircleS{29} & psycopg2.OperationalError                        &   13 &   5 & 4 \\
      \mycircleS{30} & NotImplementedError                             &   12 &   5 & 6 \\
      \mycircleS{31} & django.db.migrations.exceptions.BadMigrationError &    9 &   7 & 4 \\
      \mycircleS{32} & sphinx.errors.ExtensionError                    &    6 &   6 & 5 \\
      \bottomrule
    \end{tabular}
  }
\footnotesize
\begin{flushleft}
\textbf{Note:} Total Occ. = Total Occurrence Count; Task Prev. = Task-Level Prevalence; Agent Prev. = Agent-Level Prevalence. Sorted in descending order by Total Occ.
\end{flushleft}
\end{table}

\noindent\textbf{I. Python Built-in Errors:}
This category encompasses errors that originate from Python’s standard error/exception hierarchy. We further subdivide this category as follows:

\begin{itemize}[leftmargin=*]
  \item \textbf{I.A Dependency Errors:}
    \mycircleT{1} \textit{ModuleNotFoundError} and \mycircleT{6} \textit{ImportError} indicate that agents frequently fail to resolve and import external dependencies. This suggests a weakness in understanding or configuring the required execution environment.

  \item \textbf{I.B Parsing/Compilation Errors:}
    Errors such as \mycircleT{\small{11}} \textit{UnicodeDecodeError}, \mycircleT{\small{12}} \textit{SyntaxError}, \mycircleT{\small{18}} \textit{UnicodeEncodeError}, and \mycircleT{\small{19}} \textit{IndentationError} indicate code that does not adhere to proper syntax or encoding standards. These errors point to limitations in the agents’ internal validation of generated code.

  \item \textbf{I.C Type and Access Errors:}
    This subgroup includes \mycircleT{2} \textit{TypeError}, \mycircleT{3} \textit{AttributeError}, \mycircleT{7} \textit{ValueError}, \mycircleT{8} \textit{NameError}, \mycircleT{9} \textit{KeyError}, \mycircleT{\small{15}} \textit{IndexError}, \mycircleT{\small{21}} \textit{UnboundLocalError}, \mycircleT{\small{26}} \textit{LookupError}, and \mycircleT{\small{30}} \textit{NotImplementedError}. Their prevalence indicates that agents struggle with proper type management, variable referencing, and accessing data structures, underscoring a need for more robust error-checking during code synthesis.

 \item \textbf{I.D Operational/System Errors:}  
    Errors such as \mycircleT{\small{16}} \textit{FileNotFoundError}, \mycircleT{\small{20}} \textit{OSError}, and \mycircleT{\small{25}} \textit{subprocess.CalledProcessError} point to difficulties in managing file systems and system-level operations. These issues suggest that agents encounter challenges when interacting with external system resources.

\item \textbf{I.E Control Flow Errors:}  
\mycircleT{\small{23}} \textit{RecursionError} occurs when an agent's recursive function exceeds Python's maximum recursion depth, typically due to flawed logic.
\end{itemize}

\noindent\textbf{II. Custom-Defined Exceptions:} This category groups errors that are not part of Python’s built-in error/exception hierarchy but are defined by the repository or its associated third-party libraries. These errors and exceptions reflect challenges in adapting to domain-specific constraints. We subdivide them as follows:

\begin{itemize}[leftmargin=*]
    \item \textbf{II.A Database-Related Errors:}  
    \mycircleT{4} \lstinline[basicstyle={\itshape}]{django.db.utils.OperationalError}, \mycircleT{5} \textit{sqlite3.OperationalError}, \mycircleT{\small{10}} \textit{django.db.utils.IntegrityError}, \mycircleT{\small{13}} \textit{sqlite3.IntegrityError}, \mycircleT{\small{27}} \textit{django.db.utils.ProgrammingError}, \mycircleT{\small{28}} \textit{django.db.utils.NotSupportedError}, \mycircleT{\small{29}} \textit{psycopg2.OperationalError}, and \mycircleT{\small{31}} \textit{django.db.migrations.exceptions.BadMigrationError} capture issues directly tied to database interactions and integrity constraints. Their prevalence suggests that agents struggle with generating correct database interactions, formulating valid SQL queries, and managing schema migrations. Overall, agents appear to lack domain-specific knowledge of database programming and schema design, indicating a need for targeted training.

  \sloppy
  \item \textbf{II.B Framework Model/Validation Errors:}  \mycircleT{\small{14}} 
  \lstinline[basicstyle={\itshape\normalsize}]{django.core.exceptions.FieldError}
  and \mycircleT{\small{17}} 
  \lstinline[basicstyle={\itshape\normalsize}]{django.core.exceptions.ValidationError} 
  indicate that agents often generate code that violates the repository’s domain-specific data models or validation rules, pointing to a need for deeper familiarity with the framework’s conventions.
 
  \item \textbf{II.C Framework Management/Command Errors:} \mycircleT{\small{22}} \lstinline[basicstyle={\itshape}]{django.core.management.base.SystemCheckError}
  and \mycircleT{\small{24}} \lstinline[basicstyle={\itshape}]{django.core.management.base.CommandError} arise during the execution of management commands and system checks, implying that agents are not fully adept at interacting with the operational aspects of the framework.
  
  \item \textbf{II.D Tool/Extension Errors:} \mycircleT{\small{32}} \textit{sphinx.errors.ExtensionError} reflects errors in integrating with external documentation tools, underscoring 
  challenges in managing third-party extensions.
\end{itemize}

Our analysis demonstrates a systematic prevalence of errors across both Python built-in and custom-defined categories. High frequencies of dependency, parsing, type/access, and operational errors indicate that errors in external configuration and internal code validation are common, while frequent database and framework-related errors reveal challenges with domain-specific constraints. These findings highlight common occurrences of errors in the real-world GitHub issue resolution process, revealing current weaknesses in agents and guiding their future enhancement to avoid or handle these prevalent errors.



\begin{answer_env}{2}{Agents frequently encounter dependency, parsing, type/access, and operational errors during agents github issue solving process, highlighting issues in external configurations and internal validation. Additionally, frequent database and framework errors suggest that agents often lack the domain-specific knowledge required to handle specialized environments.}
\end{answer_env}


%% file: sections/solving_phase_challenging_error_analysis.tex
\section{Challenging Error Analysis}
\label{sec:challenging_errors_analysis}
\subsection{Experimental Setup}

To address 
``\textbf{RQ3:} \textit{Which error types are particularly challenging for software development agents to recover from during the GitHub Issue solving process?}'' 
we define \textit{challenging errors} as those that \emph{recur} within a task. In other words, if the same error appears repeatedly during the resolution phase of an agent, it suggests that the agent fails to resolve the error on its initial encounter or does not learn from it, leading to persistent recurrence. This recurrence serves as an indicator that the error is particularly challenging for the agent.

We propose the following metrics to quantify challenging errors, which are selected based on the following criteria:
\textbf{1) RQ-alignment} – every metric traces back to the information demanded by our research questions (e.g., recurrence ratio gauges error persistence);
\textbf{2) Complementarity} – Frequency, scope, and recurrence metrics together cover distinct error facets, preventing reliance on any one indicator;
\textbf{3) Practicality} – the metrics should be lightweight, interpretable, and suitable for process-level error analysis of agents in other software-engineering tasks.


\begin{table*}[t]
  \caption{Challenging Errors}
  \label{tab:agg_challenging_errors}
  \resizebox{\textwidth}{!}{%
    \begin{tabular}{r l c c o c c c c}
      \toprule
      \textbf{Rank} & \textbf{Error Type} & \textbf{Uni. Err. Inst.} & \textbf{Rec. Err. Inst.} & \textbf{\textcolor{deeporange}{Rec. Ratio $\downarrow$}} & \textbf{Avg. Rec. Cnt.} & \textbf{Max. Rec. Cnt.} & \textbf{Affected Ts.} & \textbf{Affected Ags.} \\
      \midrule
       1  & OSError                              & 7   & 5   & 71.43\% & 5.80  & 11  & 3  & 4 \\
       2  & django.db.utils.IntegrityError         & 26  & 15  & 57.69\% & 10.47 & 29  & 7  & 6 \\
       3  & IndentationError                     & 18  & 9   & 50.00\% & 2.44  & 4   & 7  & 5 \\
       4  & sqlite3.IntegrityError                 & 18  & 9   & 50.00\% & 12.22 & 29  & 6  & 5 \\
       5  & django.core.exceptions.FieldError      & 60  & 23  & 38.33\% & 3.04  & 8   & 14 & 7 \\
       6  & SyntaxError                          & 61  & 21  & 34.43\% & 4.95  & 21  & 21 & 5 \\
       7  & IndexError                           & 58  & 16  & 27.59\% & 2.81  & 7   & 8  & 6 \\
       8  & sqlite3.OperationalError             & 302 & 81  & 26.82\% & 3.47  & 31  & 49 & 7 \\
       9  & django.db.utils.OperationalError       & 323 & 86  & 26.63\% & 3.37  & 30  & 52 & 7 \\
      10  & KeyError                             & 146 & 31  & 21.23\% & 3.03  & 15  & 17 & 8 \\
      11  & UnicodeEncodeError                   & 32  & 6   & 18.75\% & 2.17  & 3   & 3  & 4 \\
      12  & ImportError                          & 337 & 62  & 18.40\% & 3.02  & 11  & 55 & 8 \\
      13  & TypeError                            & 513 & 93  & 18.13\% & 6.11  & 106 & 66 & 7 \\
      14  & ValueError                           & 300 & 49  & 16.33\% & 3.78  & 26  & 30 & 8 \\
      15  & NameError                            & 176 & 28  & 15.91\% & 3.86  & 34  & 22 & 8 \\
      16  & ModuleNotFoundError                  & 785 & 110 & 14.01\% & 3.44  & 43  & 80 & 8 \\
      17  & AttributeError                       & 611 & 84  & 13.75\% & 3.74  & 31  & 63 & 8 \\
      \bottomrule
    \end{tabular}%
  }
  \footnotesize
  \begin{flushleft}
  \textbf{Note:} \textbf{Uni. Err. Inst.} = Unique Error Instances; \textbf{Rec. Err. Inst.} = Recurred Error Instances; \textbf{Rec. Ratio} = Recurrence Ratio (Rep. Err. Inst. / Uni. Err. Inst.); \textbf{Avg. Rec. Cnt.} = Average Recurrence Count; \textbf{Max. Rec. Cnt} = Maximum Recurrence Count; \textbf{Affected Ts.} = Number of tasks where the error recurred; \textbf{Affected Ags.} = Number of agents that encountered recurring instances of the error.
  \end{flushleft}
\end{table*}

\begin{itemize}[leftmargin=*]
    \item \textbf{Unique Error Instances (UEI):} The count of distinct error occurrences extracted from the solving phase trajectories. Each unique instance is defined by a unique combination of the agent, the task, and the specific error line.

    \item \textbf{Recurred Error Instances (REI):} The count of Unique Error Instances (UEI) that occur more than once within a task.

    \item \textbf{Recurrence Ratio (RR):} The fraction of error instances that recur, defined as 

    \begin{equation}
    Recurrence\ Ratio (RR) = \frac{REI}{UEI}
    \end{equation}
    
    A higher Recurrence Ratio indicates that a larger proportion of error instances recur within a task, suggesting that when an agent encounters this type of error, it is typically challenging to resolve and tends to reoccur during the task-solving process.

    \item \textbf{Total Recurrence Count (TRC):} The sum of all repeated occurrences of the Recurred Error Instances (REI) for a given error type, representing the overall volume of recurrences.

    \item \textbf{Average Recurrence Count (ARC):} The average number of repetitions per recurring instance, reflecting the typical number of times an error recurs in a task, computed as 

    \begin{equation}
    Average\ Recurrence\ Count (ARC) = \frac{TRC}{REI}
    \end{equation}
  
  \item \textbf{Maximum Recurrence Count (MRC):} The highest number of repetitions observed for a single recurring error instance, capturing the worst-case scenario.
  
  \item \textbf{Affected Tasks (AT):} The number of unique tasks ("task" refer to a github issue) in which the error recurred (i.e., tasks where at least one agent encountered repeated instances of that error).
  
  \item \textbf{Affected Agents (AA):} The number of agents that encountered recurring instances of the error (i.e., agents for which the error recurred in at least one task).
\end{itemize}

These metrics provide a nuanced view of error persistence. The \textit{Recurrence Ratio} quantifies the fraction of error instances that recur, while the frequency metrics (i.e., $TRC$, $ARC$, and $MRC$) indicate the overall severity of these recurrences. Additionally, the \textit{Affected Tasks} and \textit{Affected Agents} metrics reveal how widely an error is distributed across tasks and agents. Together, these metrics form the foundation of our analyses for RQ3.

\subsection{Experimental Results}

Table~\ref{tab:agg_challenging_errors} presents the aggregated challenging error metrics computed across all agents from the solving process trajectories. 
This table ranks the error types by their \textit{Recurrence Ratio}s; a higher ratio indicates an error type that is more difficult for agents to resolve and tends to reoccur during the task-solving phase.
The table also reports the number of unique tasks (across agents) and the number of agents that encountered each error type, along with the \textit{Average Recurrence Count} and \textit{Maximum Recurrence Count}.

 Among the 32 unique error types identified in the solving process trajectories (see Table~\ref{tab:agg_challenging_errors}), 15 error types (e.g., \textit{UnicodeDecodeError} and \textit{FileNotFoundError}) were prevalent but did not recur within individual tasks. This indicates that agents typically resolve these errors on their first occurrence or that they are not triggered in subsequent steps, and thus they were excluded from our aggregated challenging error metrics. In contrast, the remaining error types recurred within tasks, signifying errors that persistently hinder agents from resolving and generating effective patches. This filtering 
helps to identify areas where enhanced error-handling strategies are required. 


\begin{itemize}[leftmargin=*]
    \item \textbf{System Operation Errors:} Although repeated occurrences of \textit{OSError} (Rank 1) are observed in only 4 agents across 3 tasks, it exhibits the highest Recurrence Ratio (71.43\%) with an Average Recurrence Count of 5.80 and a Maximum Recurrence Count of 11. This indicates that when \textit{OSError} occurs, it tends to persist, likely due to issues with file system interactions or system calls, posing significant challenges for agents to resolve.

  \item \textbf{Database-Related Errors:} Both \textit{django.db.utils.IntegrityError} (Rank 2) and \textit{sqlite3.IntegrityError} (Rank 4) display high Recurrence Ratios (approximately 57.69\% and 50.00\%, respectively) along with significant Average Recurrence Counts (10.47 and 12.22, respectively). This indicates that agents face notable difficulties with database integrity issues, likely due to challenges in formulating correct SQL queries or handling schema constraints.
  
  \item \textbf{Syntax and Indentation Errors:} The Recurrence Ratios for \textit{SyntaxError} (Rank 6) and \textit{IndentationError} (Rank 3) are 34.43\% and 50.00\%, respectively, indicating that even advanced agents sometimes struggle with Python code parsing, although it might be expected to resolve such errors easily.
  In contrast, \textit{ModuleNotFoundError} (Rank 16) is prevalent in the overall solving process (as shown in RQ2) but exhibits a much lower Recurrence Ratio (14.01\%), suggesting that while such module-missing errors occur frequently, agents are often able to resolve them effectively.
  
  \item \textbf{Other Built-in Errors:} Errors such as \textit{TypeError} (Rank 13), \textit{NameError} (Rank 15), and \textit{AttributeError} (Rank 17) exhibit lower Recurrence Ratios (ranging from 13.75\% to 18.13\%), suggesting that although these errors are common, agents can generally recover from these errors.
  
\end{itemize}

\begin{answer_env}{3}{Among the 32 prevalent error types, 17 show recurrence within a single task resolution process, indicating they are challenging to fix. System errors like \textit{OSError} exhibit high recurrence, and database errors, such as \textit{django.db.utils.IntegrityError}, reoccur frequently, highlighting agents' struggles with database integrity. In contrast, although \textit{ModuleNotFoundError} is common, its low recurrence ratio indicates that agents can successfully fix most instances. Overall, these recurring errors pose significant challenges and warrant targeted improvements.}
\end{answer_env}

\subsection{Supplementary Error Severity Analysis}
\label{sec:severity_analysis}

To assess the severity of each \emph{challenging error} in Table~\ref{tab:agg_challenging_errors}, we compute how often it coincides with an \emph{unsuccessful} patch. Specifically, for each error we gather every \textit{(agent, task instance)} pair that triggers it and record whether the patched code builds successfully and passes the test cases. The ratio of failed pairs to total pairs—\textbf{failure rate}—serves as a severity proxy. Because a task instance can surface multiple errors, this association is indicative rather than causal.

{
\begin{table}[ht]

  \caption{Failure Statistics of Errors}
  \label{tab:failure_challenging_errors}
  \centering
  \resizebox{\columnwidth}{!}{%
    \begin{tabular}{r l c c o}
      \toprule
      \textbf{Rank} & \textbf{Error Type} & \textbf{Tot. Pairs} & \textbf{Fail. Pairs} & \textbf{\textcolor{deeporange}{Fail. Rate $\downarrow$}} \\
      \midrule
       1  & OSError                              &   5 &   4 & 80.00\% \\
       2  & sqlite3.IntegrityError               &   9 &   7 & 77.78\% \\
       3  & IndentationError                     &   7 &   5 & 71.43\% \\
       4  & django.db.utils.IntegrityError       &  10 &   7 & 70.00\% \\
       5  & KeyError                             &  24 &  16 & 66.67\% \\
       6  & django.db.utils.OperationalError     &  78 &  48 & 61.54\% \\
       7  & NameError                            &  24 &  14 & 58.33\% \\
       8  & sqlite3.OperationalError             &  74 &  43 & 58.11\% \\
       9  & AttributeError                       &  78 &  45 & 57.69\% \\
      10  & ValueError                           &  46 &  26 & 56.52\% \\
      11  & TypeError                            &  85 &  47 & 55.29\% \\
      12  & ModuleNotFoundError                  & 104 &  55 & 52.88\% \\
      13  & SyntaxError                          &  21 &  11 & 52.38\% \\
      14  & django.core.exceptions.FieldError    &  20 &  10 & 50.00\% \\
      15  & ImportError                          &  55 &  23 & 41.82\% \\
      16  & UnicodeEncodeError                   &   5 &   2 & 40.00\% \\
      17  & IndexError                           &  15 &   3 & 20.00\% \\
      \bottomrule
    \end{tabular}%
  }
  
  \begin{flushleft}
  \footnotesize
  \textbf{Note:} \textbf{Tot. Pairs} = number of \textit{(agent, task instance)} pairs that triggered the error; 
  \textbf{Fail. Pairs} = pairs whose final solution failed; 
  \textbf{Fail. Rate} = \textbf{Fail. Pairs} / \textbf{Tot. Pairs}. 
  \end{flushleft}

\end{table}
}

Results in Table~\ref{tab:failure_challenging_errors} reveal a clear severity hierarchy. Environment-level failures dominate: \texttt{OSError} and database-integrity errors result in failure rates of approximately 75\%. Syntax and name-resolution errors follow with rates around 60\%, while logic-level issues such as \texttt{IndexError} exhibit a lower failure rate near 20\%. This suggests that agents are more capable of addressing intra-file logic bugs than external or environment-related faults.

\begin{table*}[t!]
  \caption{Categorization of Task Failures}
  \label{tab:unresolved_summary}
  \resizebox{\textwidth}{!}{%
    \begin{tabular}{l c  cc  cccc}
      \toprule
      \textbf{Agent} & \textbf{Unresolved Tasks} & \multicolumn{2}{c}{\textbf{Patch Failures}} & \multicolumn{4}{c}{\textbf{Testing Failures}} \\
      \cmidrule(lr){3-4} \cmidrule(lr){5-8}
                   &                          & \textbf{No Patch Generated} & \textbf{Patch Failed to Apply} & \textbf{No Test Log} & \textbf{Contains Parsing/Runtime Errors} & \textbf{Contains AssertionErrors Only} & \textbf{No Explicit Errors Extracted} \\
      \midrule
      W\&B Programmer          & 177 & 1  & 2 & 0 & 147 & 21 & 6 \\
      Blackbox AI              & 186 & 20 & 1 & 0 & 146 & 15 & 4 \\
      Devlo                    & 209 & 0  & 0 & 0 & 182 & 24 & 3 \\
      CodeAct                  & 235 & 6  & 0 & 1 & 199 & 24 & 5 \\
      Engine Labs              & 241 & 8  & 0 & 0 & 191 & 22 & 20 \\
      MarsCode                 & 250 & 2  & 0 & 0 & 216 & 27 & 5 \\
      Tools                    & 255 & 14 & 0 & 3 & 203 & 29 & 6 \\
      CodeShell                & 279 & 13 & 0 & 1 & 237 & 21 & 7 \\
      \bottomrule
    \end{tabular}%
  }
\end{table*}

%% file: sections/testing_phase_error_analysis.tex

\section{Failure and Cross-Phase Error Analysis}
\label{sec:patch_failure}

To answer ``\textbf{RQ4:} \textit{What are the underlying failure reasons for patches, and which errors encountered during the testing phase were previously observed (but not resolved) during the solving phase?}'', our analysis is divided into three parts, including an examination of unresolved tasks with different failure reasons, an investigation of cross-phase errors, and an analysis of failures for which no explicit Python execution errors were extracted.

\subsection{Categorization of Task Failures}
\subsubsection{Experimental Design}

Our experimental design comprises three main steps to investigate the underlying reasons for patch failures in resolving GitHub issues:

\begin{enumerate}[leftmargin=*, label=\arabic*.]
  \item \textbf{Counting Unresolved Tasks.}  
  We extract the total number of unresolved tasks for each agent from the result report file.
  
  \item \textbf{Categorizing Failure Reasons.}  
  To understand why tasks remain unresolved, we classify failures into two main groups:
  \begin{itemize}[leftmargin=*]
    \item \textbf{Patch Failures.}  
    \begin{itemize}[leftmargin=*]
      \item \textit{No Patch Generated} --- the agent failed to generate a patch.
      \item \textit{Patch Failed to Apply} --- a patch was generated but could not be applied (e.g., due to syntax errors or patch conflicts).
    \end{itemize}
    \item \textbf{Testing Failures.}  
    For tasks where a patch was applied, we classify failures based on testing outputs:
    \begin{itemize}[leftmargin=*]
      \item \textit{Parsing/Runtime Errors} --- the patched code is not executable due to errors.
      \item \textit{Assertion Errors Only} --- the patched code runs but fails to meet its intended functionality.
    \end{itemize}
  \end{itemize}
  
  \item \textbf{Integration and Analysis.}  
  Finally, we aggregate the counts from the above to determine the overall number of unresolved tasks per agent and to analyze the distribution of failure reasons.
\end{enumerate}

\subsubsection{Experiment Result}

Our analysis of unresolved tasks, summarized in Table~\ref{tab:unresolved_summary}, reveals that most unresolved GitHub issues failed primarily due to Python execution failures during testing, often stemming from parsing or runtime errors. In contrast, tasks exhibiting only assertion errors suggest that although the patch is executable, it does not fully address the underlying issue. Additionally, unresolved tasks with no explicit Python error messages warrant further manual investigation.

These failures indicate that agents misjudge the effectiveness of their patches and end their solving phase prematurely with an incorrect assessment and thus submit patches that still contain critical bugs and unresolved issues.
This observation aligns with existing studies~\cite{zheng2023judging,fu2023gptscore,chen2024mllm,zhuge2025agentasajudge}, that also highlight agents' limitations such as positional/verbosity bias, hallucinatory responses, and inconsistencies in their judgment,
while recognizing that LLMs/agents can approach human-level judgment.

\subsection{Analysis of Cross-Phase Error Pairs}

\subsubsection{Experimental Setup}We aggregate error pairs extracted from the solving and testing logs, referred to as \textit{cross-phase error pairs}. These pairs represent instances where an error observed during the solving phase reoccurs during testing, suggesting that these errors are stealthy and fail to be adequately detected and resolved by agents before patch submission.

\begin{table}[h]
\caption{Cross-Phase Error Pairs}
\label{tab:cross_phase_error_summary}
\centering
\resizebox{\columnwidth}{!}{%
\begin{tabular}{lccc}
\toprule
\textbf{Error Type} & \textbf{Pairs} & \textbf{Num. Agents} & \textbf{Num. Tasks} \\
\midrule
TypeError                                & 10 & 4 & 9 \\
ModuleNotFoundError                      & 9  & 8 & 2 \\
AttributeError                           & 6  & 5 & 3 \\
NameError                                & 6  & 3 & 6 \\
django.core.exceptions.ValidationError   & 6  & 1 & 1 \\
django.db.utils.IntegrityError             & 5  & 1 & 1 \\
ValueError                               & 4  & 2 & 2 \\
ImportError                              & 1  & 1 & 1 \\
RecursionError                           & 1  & 1 & 1 \\
SyntaxError                              & 1  & 1 & 1 \\
UnboundLocalError                        & 1  & 1 & 1 \\
sympy.polys.polyerrors.PolynomialError     & 1  & 1 & 1 \\
\bottomrule
\end{tabular}%
}
\footnotesize
\begin{flushleft}
\textbf{Note:} ``Number of Pairs'' is the total count of matching error pairs; ``Number of Agents'' is the number of unique agents that encountered the pair; and ``Number of Tasks'' is the number of unique tasks in which the pair occurred.
\end{flushleft}
\end{table}

\subsubsection{Experiment Results}
Table~\ref{tab:cross_phase_error_summary} shows that the most frequently observed cross-phase error pair is \textit{TypeError}, with 10 pairs occurring across 4 agents and 9 tasks. In contrast, although \textit{ModuleNotFoundError} is encountered by 8 agents, it appears in only 2 tasks, indicating that while many agents experience this error, its occurrence is confined to a limited number of cases. Similarly, \textit{AttributeError} and \textit{NameError} display moderate prevalence, whereas the remaining error types occur only sporadically. Overall, these findings highlight that certain error types are stealthy, persisting from the solving phase to the testing phase, and thereby are potential targets for improving error detection and recovery mechanisms.

\subsection{Manual Investigation of Failures}
\subsubsection{Experimental Design}
We manually reviewed 56 unresolved task cases that had no explicit execution errors extracted. For each task, we first examined the \texttt{report.json} file to identify failed unit tests, and then inspected the corresponding \lstinline{testing execution output.txt} logs to determine the specific failure reasons.

\begin{table}[ht]
\caption{Implicit Task Failures}
\label{tab:case_count}
\centering
\resizebox{\columnwidth}{!}{%
  \begin{tabular}{lll}
    \toprule
    \multicolumn{3}{c}{\textbf{Tests Not Executed}} \\
    \midrule
    \textbf{Reason} & \textbf{Cases} & \textbf{Explanation} \\
    Environment Setup Failure & 18 & Testing environment failed to initialize. \\
    SystemExit Error          & 16 & SystemExit interrupted execution. \\
    \midrule
    \multicolumn{3}{c}{\textbf{Tests Executed}} \\
    \midrule
    \textbf{Reason} & \textbf{Cases} & \textbf{Explanation} \\
    False Negative Reporting  & 10 & Tests passed but were reported as failures. \\
    Unexpected Error Format   & 11 & Errors displayed in an unexpected format. \\
    No Explicit Error Information & 1  & Execution failed without clear error details. \\
    \bottomrule
  \end{tabular}%
}
\end{table}

\subsubsection{Experiment Result}
We categorized these cases into two groups with five subcategories (as shown in Table~\ref{tab:case_count}). Of the 56 cases, 34 resulted from tests not being executed (18 due to environment failures and 16 due to SystemExit errors), while the remaining 22 cases involved executed tests. Among these, 10 cases were false negatives (i.e., all tests passed but were marked as unresolved), 11 cases displayed uncommon error formats (e.g., ``Failed: DID NOT RAISE <class `ValueError'>'', which appears to be an atypical presentation of an AssertionError), and 1 case lacked clear error details. 

\subsection{SWE-Bench Bugs}
Notably, in our manual investigation of tasks without explicit Python execution errors, we identified 3 tasks in which all 8 agents failed the testing phase for the same reason, suggesting potential bugs in the SWE-Bench platform. 

\begin{table}[ht]
\centering
\caption{SWE-Bench Bugs Encountered by All 8 Agents}
\label{tab:swebench_bugs}
\resizebox{\columnwidth}{!}{%
  \begin{tabular}{lll}
    \toprule
    \textbf{Task Id} & \textbf{Failure Reason} & \textbf{Status} \\
    \midrule
    astropy-7606   & Passed all test cases but falsely marked as unresolved & Fixed \\
    astropy-8707   & Failure in setting up the test cases                     & Fixing in Progress \\
    astropy-8872   & Failure in collecting the test cases                       & Fixing in Progress \\
    \bottomrule
  \end{tabular}%
}
\end{table}

As shown in Table~\ref{tab:swebench_bugs}, in task \textit{astropy-7606}, all agents passed every test case yet were falsely marked as unresolved. In tasks \textit{astropy-8707} and \textit{astropy-8872}, failures occurred during the setup and collection of test cases, respectively, preventing a fair evaluation of the patches. We submitted 3 bug reports to the SWE-Bench maintainers via GitHub. By the time of our paper submission, all the reports have received responses and the issues have been confirmed: 1 issue has also been fixed while the other 2 are being resolved.

\begin{answer_env}{4}{Our results indicate that most unresolved tasks stem from testing-phase failures (primarily parsing and runtime errors), suggesting that agents do not accurately assess patch quality. Moreover, persistent cross-phase errors (e.g., recurring TypeError) imply that some issues remain unresolved. Additionally, we identified three bugs in the SWE-Bench platform that undermine the accuracy and fairness of the evaluation for agents.}
\end{answer_env}

%% file: sections/discussion.tex
\section{Discussion}
\label{sec:discussion}
\subsection{Implication}
Our study underscores the demanding need for techniques that can facilitate agent recovery from the most prevalent and challenging errors, ultimately reducing computational overhead and promoting greener software development practices. By minimizing errors that repeatedly distract an agent, researchers can help agents produce more accurate final solutions. Additionally, there is a pressing need for a standardized format for trajectory files in software development agent workflows; currently, these trajectory logs take diverse forms (e.g., JSON, Markdown, plain text). Establishing a unified format would enable more comprehensive comparative analyses of agent behavior and streamline future evaluations of agents’ strengths and weaknesses.

\subsection{Relevance with Open-Sourced Agents}
\label{subsec:agent_relevance}

We analyse the phases and error-handling strategies of four public agents (Table \ref{tab:agent_perf}). \textbf{SWE-Agent} integrates an inline linter that blocks syntax faults (e.g., \texttt{IndentationError}), improving resolution rate by 3\%~\cite{yang2024swe}. 
\textbf{AutoCodeRover} employs a \emph{Reviewer} that runs the reproducer test on the patched code to surface new compilation or run-time failures (e.g., \texttt{TypeError}); surviving patches then pass a full regression suite that detects functional regressions \cite{zhang2024autocoderover,ruan2024specrover}. 
\textbf{Agentless} applies an analogous two-stage filter—reproduction test followed by regression tests—to ensure patches execute cleanly \cite{xia2024agentless}.
\textbf{SWE-Llama}, a pure RAG baseline, offers no error detection \cite{jimenez2024swebench}.

\begin{table}[h]
\centering
\small
\setlength{\tabcolsep}{3pt}
\renewcommand{\arraystretch}{1.12}
\caption{Agent Workflow Phases}
\label{tab:agent_perf}
\begin{tabular}{p{0.24\columnwidth}  
                p{0.42\columnwidth}  
                C{0.10\columnwidth}  
                C{0.12\columnwidth}} 
\toprule
\textbf{Agent} & \textbf{Agentic Phases Summary} & \textbf{Resolved} & \textbf{Date} \\
\midrule
\raggedright SWE-Llama 7B &
Retrieves code with BM25, wraps instructions + example diff, emits repository-wide patch. &
1.4\% & 10/Oct/23 \\
\raggedright Agentless-1.5\\(Claude-3.5 Sonnet) &
Hierarchically localizes code, samples diff patches, keeps patch passing regression + reproduction tests. &
50.8\% & 02/Dec/24 \\
\raggedright AutoCodeRover v2.1(Claude-3.5 Sonnet) &
Reproduces bug, localises context, summarises intent, patches, double-tests, selects best validated fix. &
51.6\% & 22/Jan/25 \\
\raggedright SWE-Agent\\(Claude 4 Sonnet) &
Searches, views, edits code via ACI; linter guardrail blocks syntax errors. &
66.6\% & 22/May/25 \\
\bottomrule
\end{tabular}
\footnotesize
\begin{flushleft}
\textbf{Note:} Latest score per agent reported by (8 Jul 2025) from SWE-bench Verified.
\end{flushleft}
\end{table}

\noindent
\subsection{Future Work}
\label{sec:future_work}
Our findings offer several insights and highlight urgent research challenges for future work:

\noindent\textbf{1. Error-Prone Benchmarks.}
As AI-based software development advances, there is an urgent need for benchmarks to assess their capabilities. Several benchmarks evaluate various aspects of automated software development beyond mere functionality, as shown in studies~\cite{tony2023llmseceval,fu2024constrained,siddiq2022securityeval} that include vulnerability-prone scenarios. These benchmarks determine whether generated code meets functional requirements without introducing vulnerabilities. Therefore, developing specialized benchmarks for error-prone scenarios, such as database integrity challenges and missing dependencies, could improve our understanding of agent error handling, reveal their strategies and limitations.

\noindent\textbf{2. Proactive Error Avoidance.}
Improving the workflow of software development agents is a promising direction~\cite{terragni2025future}. Enhancing an agent’s process with early, proactive measures that preempt common yet pervasive errors may reduce overhead and improve resolution rates. For instance, implementing preemptive dependency checks~\cite{horton2019dockerizeme,jin2024pyanalyzer,ye2022knowledge} and incorporating conflict monitoring steps~\cite{wang2020watchman} can help prevent errors such as \textit{ModuleNotFoundError}. Additionally, using static analysis tools~\cite{vassallo2020developers} further enhances early error detection: MyPy can identify type-related issues (e.g., \textit{TypeError}), while tools like Pylint and Flake8 can catch syntax errors, indentation errors, and other code style violations. This approach not only reduces wasted execution time and resources but also minimizes distractions for agents, allowing them to focus on their primary tasks before patch submission.

\noindent\textbf{3. Integrated Error Recovery.}
Incorporating effective error recovery strategies~\cite{randell2025looking} directly into agents—via retrieval-augmented generation (RAG) approaches~\cite{shi2024efficient} or by integrating established error detection and resolution methods—could significantly improve their ability to address challenging issues. There are several existing repair techniques for errors, for example, for \textit{TypeError}\cite{peng2024domain,chow2024pyty,oh2024towards}, dependency-related errors\cite{mukherjee2021fixing, wang2022smartpip,ye2022knowledge}, and \textit{SyntaxError}~\cite{gore2023syntax,santos2018syntax}. How to integrate these existing techniques into the agentic workflow and ensure effective collaboration between the tool and the agent is a promising direction.

\noindent\textbf{4. Greener AI-Driven Software Development.}
Efficient and sustainable large language models for software engineering (LLM4SE) are critical for the future of the field~\cite{cruz2024innovating}. By quantifying the energy costs associated with repeated debugging cycles and prolonged error-resolution phases, we can pave the way for more sustainable practices. Future work should focus on developing effective and accurate measurement methods and metrics to better quantify the energy cost in the problem-solving process, and on identifying optimization strategies that improve agent efficiency~\cite{shi2024efficient}. This could involve reducing the computational resources used for repairing code errors without compromising performance.

\noindent\textbf{5. Cross-Benchmark Exploration.} We study agents in GitHub-issue resolution tasks. Other benchmarks—BigCodeBench (function synthesis)~\cite{zhuo2025bigcodebench}, EvoCodeBench (code evolution)~\cite{li2024evocodebench}, and RepoBench (repository completion)~\cite{liu2024repobench}—cover different contexts but presently publish no agent-run data, so we cannot replicate our error analysis on them yet. When richer traces emerge, future work should test whether our failure patterns persist across these tasks.
  
\subsection{Threats to Validity}
\noindent\textit{\textbf{Threats to Internal Validity.}}
Internal validity concerns whether our findings accurately reflect the causal relationships in our study. One potential bias arises from selecting only eight agents with high-quality trajectories, which may exclude agents with different error-handling behaviors. Our data extraction was based on parsing \textit{OBSERVATION} sections and test logs, which may miss or misidentify errors if formatting is inconsistent. In addition, relying on the stability of the SWE-Bench Verified dataset means that undisclosed bugs or changes could confound our results. To mitigate these threats, we validated error extraction through manual checks, refined and standardized our data collection process, and reported identified platform bugs to the SWE-Bench maintainers.

\smallskip
\noindent\textit{\textbf{Threats to External Validity.}}
External validity concerns the generalizability of our findings beyond this study. Although SWE-Bench covers 500 real GitHub issues from 12 Python repositories, these may not represent all software engineering tasks or domains, and our conclusions may not extend to other languages or environments. Moreover, our selected top-performing agents may not capture the full variability of broader agent populations or future advancements. To mitigate these threats, we chose SWE-Bench for its relative realism; the tasks were validated by software engineers with support from the OpenAI preparedness team, ensuring that this high-quality benchmark accurately reflects real-world GitHub tasks. We also documented our selection criteria for future replication or comparison as more comprehensive benchmarks emerge.

\smallskip
\noindent\textit{\textbf{Threats to Construct Validity.}}
Construct validity concerns whether our metrics accurately capture the intended error characteristics. One potential threat is that our metrics may not fully capture the nuanced severity or context of each error. To mitigate this, for both prevalent error analysis and challenging error analysis, we adopted multiple complementary metrics. For example, the Occurrence Ratio measures the proportion of an error's recurrence tendency, while Average Repetition Count and Maximum Repetition Count indicate the severity and persistence of those errors. Additionally, Affected Tasks and Affected Agents capture the overall prevalence of the error. This multi-metric approach ensures a more robust and comprehensive assessment.

%% file: sections/related_work.tex
\section{Related Work}
\label{sec:related_work}
Several studies have evaluated code solutions generated by large language models (LLMs) and software development agents. For example, research by Pearce et al.~\cite{pearce2022asleep} and Majdinasab et al.~\cite{majdinasab2024assessing} has revealed significant security vulnerabilities in Copilot-generated code, while studies by Asare et al.~\cite{asare2023github} and Hamer et al.~\cite{hamer2024just} have compared LLM-generated code with human-written code, noting that LLMs can sometimes perform comparably despite introducing certain vulnerabilities. Other works, such as those by Nguyen et al.~\cite{nguyen2022empirical}, Liu et al.~\cite{liu2024no, liu2024refining}, Rabbi et al.~\cite{rabbi2024ai}, and Siddiq et al.~\cite{siddiq2024quality}, have focused on aspects of code quality, correctness, maintainability, and complexity across various tasks. Additionally, Chen et al.~\cite{chen2024evaluating} have conducted code quality analyses on patched code to assess the current capabilities of software development agents in addressing real-world GitHub issues.

Our work differs from previous studies in two key ways. While existing research focuses on evaluating the final code solutions produced by LLMs or agents on specific coding tasks to assess their capabilities and shortcomings, our study instead focuses on the process data generated during the resolution of real-world GitHub issues—specifically, the agents' reasoning traces, execution logs, and test outputs. We believe that these artifacts offer crucial insights into how agents approach complex software engineering tasks. Moreover, rather than relying solely on static code quality metrics such as bug counts, vulnerability statistics, or complexity measures, we analyze the code errors that arise during the agents' iterative try-and-error process. This approach enables us to identify which error types agents frequently encounter and find particularly challenging, thereby highlighting areas where improvements in error avoidance and handling are urgently needed.

%% file: sections/conclusion.tex
\section{Conclusion}
\label{sec:conclusion}
We are the first to analyze errors arising from agents’ resolution process data as they solve real-world GitHub issues. In this study, we examined 3,977 solving-phase trajectories and 3,931 testing logs from eight top-ranked agents tackling 500 GitHub issues across 12 Python repositories in the SWE-Bench Verified benchmark, thereby revealing agents' current capabilities and limitations. Through a comprehensive analysis guided by four research questions, \textit{RQ1} shows that unexpected Python execution errors during task resolution correlate with lower resolution rates and increased reasoning overhead, underscoring the need for more effective error-handling strategies. \textit{RQ2} identifies prevalent errors—such as 1,053 instances of \textit{ModuleNotFoundError} and 992 instances of \textit{TypeError}—highlighting dependency and type-checking as major challenges. In \textit{RQ3}, we find that errors frequently recurring during task resolution—such as system operational errors (\textit{OSError}) and database failures (e.g., \textit{IntegrityError})—indicate that these issues are particularly challenging for agents. Lastly, \textit{RQ4} examines the reasons behind unresolved tasks during testing and uncovers three SWE-Bench platform bugs that compromise the benchmark’s correctness and fairness. These bugs have been reported and confirmed by the SWE-Bench authors.

In conclusion, our study employs a process-oriented error analysis to examine the current error behavior of software development agents and provide insights that can guide future improvements. Nevertheless, our findings are limited by the diverse data formats in use, which complicate data extraction and impede comprehensive comparisons of agent behavior. This highlights the need for a unified format that facilitates more comprehensive cross-agent analysis and streamlines future evaluations of agents’ strengths and weaknesses. As discussed in Section~\ref{sec:future_work}, future work can develop specialized error-prone benchmarks targeting scenarios such as database integrity challenges and missing dependencies, in order to deepen our understanding of agent error handling and evaluate the efficiency of automated error recovery. We should also enhance agents' error handling capabilities by implementing proactive error avoidance measures at the early stages of the agent workflow and incorporating existing knowledge and techniques to improve error recovery. Furthermore, quantifying and optimizing computational and energy costs associated with repeated and prolonged error-resolution phases is critical for building greener and more effective software development agents.


%% file: main.bib
@inproceedings{pearce2022asleep,
  title={Asleep at the keyboard? assessing the security of github copilot’s code contributions},
  author={Pearce, Hammond and Ahmad, Baleegh and Tan, Benjamin and Dolan-Gavitt, Brendan and Karri, Ramesh},
  booktitle={2022 IEEE Symposium on Security and Privacy (SP)},
  pages={754--768},
  year={2022},
  organization={IEEE}
}

@inproceedings{majdinasab2024assessing,
  title={Assessing the Security of GitHub Copilot's Generated Code-A Targeted Replication Study},
  author={Majdinasab, Vahid and Bishop, Michael Joshua and Rasheed, Shawn and Moradidakhel, Arghavan and Tahir, Amjed and Khomh, Foutse},
  booktitle={2024 IEEE International Conference on Software Analysis, Evolution and Reengineering (SANER)},
  pages={435--444},
  year={2024},
  organization={IEEE}
}

@inproceedings{nguyen2022empirical,
  title={An empirical evaluation of GitHub copilot's code suggestions},
  author={Nguyen, Nhan and Nadi, Sarah},
  booktitle={Proceedings of the 19th International Conference on Mining Software Repositories},
  pages={1--5},
  year={2022}
}

@article{asare2023github,
  title={Is github’s copilot as bad as humans at introducing vulnerabilities in code?},
  author={Asare, Owura and Nagappan, Meiyappan and Asokan, N},
  journal={Empirical Software Engineering},
  volume={28},
  number={6},
  pages={129},
  year={2023},
  publisher={Springer}
}

@inproceedings{hamer2024just,
  title={Just another copy and paste? Comparing the security vulnerabilities of ChatGPT generated code and StackOverflow answers},
  author={Hamer, Sivana and d’Amorim, Marcelo and Williams, Laurie},
  booktitle={2024 IEEE Security and Privacy Workshops (SPW)},
  pages={87--94},
  year={2024},
  organization={IEEE}
}

@article{liu2024no,
  title={No need to lift a finger anymore? assessing the quality of code generation by chatgpt},
  author={Liu, Zhijie and Tang, Yutian and Luo, Xiapu and Zhou, Yuming and Zhang, Liang Feng},
  journal={IEEE Transactions on Software Engineering},
  year={2024},
  publisher={IEEE}
}

@inproceedings{rabbi2024ai,
  title={AI writes, we analyze: The ChatGPT python code saga},
  author={Rabbi, Md Fazle and Champa, Arifa Islam and Zibran, Minhaz F and Islam, Md Rakibul},
  booktitle={Proceedings of the 21st International Conference on Mining Software Repositories},
  pages={177--181},
  year={2024}
}

@inproceedings{siddiq2024quality,
  title={Quality Assessment of ChatGPT Generated Code and their Use by Developers},
  author={Siddiq, Mohammed Latif and Roney, Lindsay and Zhang, Jiahao and Santos, Joanna Cecilia Da Silva},
  booktitle={Proceedings of the 21st International Conference on Mining Software Repositories},
  pages={152--156},
  year={2024}
}

@article{liu2024refining,
  title={Refining chatgpt-generated code: Characterizing and mitigating code quality issues},
  author={Liu, Yue and Le-Cong, Thanh and Widyasari, Ratnadira and Tantithamthavorn, Chakkrit and Li, Li and Le, Xuan-Bach D and Lo, David},
  journal={ACM Transactions on Software Engineering and Methodology},
  volume={33},
  number={5},
  pages={1--26},
  year={2024},
  publisher={ACM New York, NY}
}

@inproceedings{chen2024evaluating,
  title={Evaluating Software Development Agents: Patch Patterns, Code Quality, and Issue Complexity in Real-World GitHub Scenarios},
  author={Chen, Zhi and Jiang, Lingxiao},
  booktitle={32nd IEEE International Conference on Software Analysis, Evolution, and Reengineering (SANER 2025)},
  url={arXiv preprint arXiv:2410.12468},
  year={2024}
}

@inproceedings{jimenez2024swebench,
title={{SWE}-bench: Can Language Models Resolve Real-world Github Issues?},
author={Carlos E Jimenez and John Yang and Alexander Wettig and Shunyu Yao and Kexin Pei and Ofir Press and Karthik R Narasimhan},
booktitle={The Twelfth International Conference on Learning Representations},
year={2024},
url={https://openreview.net/forum?id=VTF8yNQM66}
}

@article{he2024llm,
  title={LLM-Based Multi-Agent Systems for Software Engineering: Literature Review, Vision and the Road Ahead},
  author={He, Junda and Treude, Christoph and Lo, David},
  journal={ACM Transactions on Software Engineering and Methodology},
  year={2024},
  publisher={ACM New York, NY}
}

@article{marjanov2022machine,
  title={Machine learning for source code vulnerability detection: What works and what isn’t there yet},
  author={Marjanov, Tina and Pashchenko, Ivan and Massacci, Fabio},
  journal={IEEE Security \& Privacy},
  volume={20},
  number={5},
  pages={60--76},
  year={2022},
  publisher={IEEE}
}

@inproceedings{kharkar2022learning,
  title={Learning to reduce false positives in analytic bug detectors},
  author={Kharkar, Anant and Moghaddam, Roshanak Zilouchian and Jin, Matthew and Liu, Xiaoyu and Shi, Xin and Clement, Colin and Sundaresan, Neel},
  booktitle={Proceedings of the 44th International Conference on Software Engineering},
  pages={1307--1316},
  year={2022}
}

@inproceedings{li2020improving,
  title={Improving bug detection and fixing via code representation learning},
  author={Li, Yi},
  booktitle={Proceedings of the ACM/IEEE 42nd International Conference on Software Engineering: Companion Proceedings},
  pages={137--139},
  year={2020}
}

@inproceedings{li2020dlfix,
  title={Dlfix: Context-based code transformation learning for automated program repair},
  author={Li, Yi and Wang, Shaohua and Nguyen, Tien N},
  booktitle={Proceedings of the ACM/IEEE 42nd international conference on software engineering},
  pages={602--614},
  year={2020}
}

@inproceedings{jiang2021cure,
  title={Cure: Code-aware neural machine translation for automatic program repair},
  author={Jiang, Nan and Lutellier, Thibaud and Tan, Lin},
  booktitle={2021 IEEE/ACM 43rd International Conference on Software Engineering (ICSE)},
  pages={1161--1173},
  year={2021},
  organization={IEEE}
}

@article{chen2019sequencer,
  title={Sequencer: Sequence-to-sequence learning for end-to-end program repair},
  author={Chen, Zimin and Kommrusch, Steve and Tufano, Michele and Pouchet, Louis-No{\"e}l and Poshyvanyk, Denys and Monperrus, Martin},
  journal={IEEE Transactions on Software Engineering},
  volume={47},
  number={9},
  pages={1943--1959},
  year={2019},
  publisher={IEEE}
}

@article{wang2021codet5,
  title={Codet5: Identifier-aware unified pre-trained encoder-decoder models for code understanding and generation},
  author={Wang, Yue and Wang, Weishi and Joty, Shafiq and Hoi, Steven CH},
  journal={arXiv preprint arXiv:2109.00859},
  year={2021}
}

@inproceedings{feng-etal-2020-codebert,
    title = "{C}ode{BERT}: A Pre-Trained Model for Programming and Natural Languages",
    author = "Feng, Zhangyin  and
      Guo, Daya  and
      Tang, Duyu  and
      Duan, Nan  and
      Feng, Xiaocheng  and
      Gong, Ming  and
      Shou, Linjun  and
      Qin, Bing  and
      Liu, Ting  and
      Jiang, Daxin  and
      Zhou, Ming",
    editor = "Cohn, Trevor  and
      He, Yulan  and
      Liu, Yang",
    booktitle = "Findings of the Association for Computational Linguistics: EMNLP 2020",
    month = nov,
    year = "2020",
    address = "Online",
    publisher = "Association for Computational Linguistics",
    url = "https://aclanthology.org/2020.findings-emnlp.139/",
    doi = "10.18653/v1/2020.findings-emnlp.139",
    pages = "1536--1547",
}

@article{guo2025deepseek,
  title={Deepseek-r1: Incentivizing reasoning capability in llms via reinforcement learning},
  author={Guo, Daya and Yang, Dejian and Zhang, Haowei and Song, Junxiao and Zhang, Ruoyu and Xu, Runxin and Zhu, Qihao and Ma, Shirong and Wang, Peiyi and Bi, Xiao and others},
  journal={arXiv preprint arXiv:2501.12948},
  year={2025}
}

@article{grattafiori2024llama,
  title={The llama 3 herd of models},
  author={Grattafiori, Aaron and Dubey, Abhimanyu and Jauhri, Abhinav and Pandey, Abhinav and Kadian, Abhishek and Al-Dahle, Ahmad and Letman, Aiesha and Mathur, Akhil and Schelten, Alan and Vaughan, Alex and others},
  journal={arXiv preprint arXiv:2407.21783},
  year={2024}
}

@article{achiam2023gpt,
  title={Gpt-4 technical report},
  author={Achiam, Josh and Adler, Steven and Agarwal, Sandhini and Ahmad, Lama and Akkaya, Ilge and Aleman, Florencia Leoni and Almeida, Diogo and Altenschmidt, Janko and Altman, Sam and Anadkat, Shyamal and others},
  journal={arXiv preprint arXiv:2303.08774},
  year={2023}
}

@article{team2023gemini,
  title={Gemini: a family of highly capable multimodal models},
  author={Team, Gemini and Anil, Rohan and Borgeaud, Sebastian and Alayrac, Jean-Baptiste and Yu, Jiahui and Soricut, Radu and Schalkwyk, Johan and Dai, Andrew M and Hauth, Anja and Millican, Katie and others},
  journal={arXiv preprint arXiv:2312.11805},
  year={2023}
}

@article{shinn2023reflexion,
  title={Reflexion: Language agents with verbal reinforcement learning},
  author={Shinn, Noah and Cassano, Federico and Gopinath, Ashwin and Narasimhan, Karthik and Yao, Shunyu},
  journal={Advances in Neural Information Processing Systems},
  volume={36},
  pages={8634--8652},
  year={2023}
}

@inproceedings{xia2024automated,
  title={Automated program repair via conversation: Fixing 162 out of 337 bugs for \$0.42 each using ChatGPT},
  author={Xia, Chunqiu Steven and Zhang, Lingming},
  booktitle={Proceedings of the 33rd ACM SIGSOFT International Symposium on Software Testing and Analysis},
  pages={819--831},
  year={2024}
}

@inproceedings{ahmed2022few,
  title={Few-shot training llms for project-specific code-summarization},
  author={Ahmed, Toufique and Devanbu, Premkumar},
  booktitle={Proceedings of the 37th IEEE/ACM international conference on automated software engineering},
  pages={1--5},
  year={2022}
}

@inproceedings{lu2023llama,
  title={Llama-reviewer: Advancing code review automation with large language models through parameter-efficient fine-tuning},
  author={Lu, Junyi and Yu, Lei and Li, Xiaojia and Yang, Li and Zuo, Chun},
  booktitle={2023 IEEE 34th International Symposium on Software Reliability Engineering (ISSRE)},
  pages={647--658},
  year={2023},
  organization={IEEE}
}

@article{liu2023your,
  title={Is your code generated by chatgpt really correct? rigorous evaluation of large language models for code generation},
  author={Liu, Jiawei and Xia, Chunqiu Steven and Wang, Yuyao and Zhang, Lingming},
  journal={Advances in Neural Information Processing Systems},
  volume={36},
  pages={21558--21572},
  year={2023}
}

@inproceedings{zhuo2025bigcodebench,
title={BigCodeBench: Benchmarking Code Generation with Diverse Function Calls and Complex Instructions},
author={Terry Yue Zhuo and Vu Minh Chien and Jenny Chim and Han Hu and Wenhao Yu and Ratnadira Widyasari and Imam Nur Bani Yusuf and Haolan Zhan and Junda He and Indraneil Paul and Simon Brunner and Chen GONG and James Hoang and Armel Randy Zebaze and Xiaoheng Hong and Wen-Ding Li and Jean Kaddour and Ming Xu and Zhihan Zhang and Prateek Yadav and Naman Jain and Alex Gu and Zhoujun Cheng and Jiawei Liu and Qian Liu and Zijian Wang and David Lo and Binyuan Hui and Niklas Muennighoff and Daniel Fried and Xiaoning Du and Harm de Vries and Leandro Von Werra},
booktitle={The Thirteenth International Conference on Learning Representations},
year={2025},
url={https://openreview.net/forum?id=YrycTjllL0}
}

@article{ma2024understand,
  title={How to understand whole software repository?},
  author={Ma, Yingwei and Yang, Qingping and Cao, Rongyu and Li, Binhua and Huang, Fei and Li, Yongbin},
  journal={arXiv preprint arXiv:2406.01422},
  year={2024}
}

@article{yang2024swe,
  title={{SWE}-agent: Agent-computer interfaces enable automated software engineering},
  author={Yang, John and Jimenez, Carlos E. and Wettig, Alexander and Lieret, Kilian and Yao, Shunyu and Narasimhan, Karthik and Press, Ofir},
  journal={Advances in Neural Information Processing Systems},
  volume={37},
  pages={50528--50652},
  year={2024}
}

@inproceedings{wang2024openhands,
  title={Openhands: An open platform for ai software developers as generalist agents},
  author={Wang, Xingyao and Li, Boxuan and Song, Yufan and Xu, Frank F and Tang, Xiangru and Zhuge, Mingchen and Pan, Jiayi and Song, Yueqi and Li, Bowen and Singh, Jaskirat and others},
  booktitle={The Thirteenth International Conference on Learning Representations},
  year={2024}
}

@inproceedings{hongjinlearn,
  title={Learn-by-interact: A Data-Centric Framework For Self-Adaptive Agents in Realistic Environments},
  author={Hongjin, SU and Sun, Ruoxi and Yoon, Jinsung and Yin, Pengcheng and Yu, Tao and Arik, Sercan O},
  booktitle={The Thirteenth International Conference on Learning Representations}
}

@article{liu2024marscode,
  title={Marscode agent: Ai-native automated bug fixing},
  author={Liu, Yizhou and Gao, Pengfei and Wang, Xinchen and Liu, Jie and Shi, Yexuan and Zhang, Zhao and Peng, Chao},
  journal={arXiv preprint arXiv:2409.00899},
  year={2024}
}

@article{xia2024agentless,
  title={Agentless: Demystifying llm-based software engineering agents},
  author={Xia, Chunqiu Steven and Deng, Yinlin and Dunn, Soren and Zhang, Lingming},
  journal={arXiv preprint arXiv:2407.01489},
  year={2024}
}

@article{ma2024lingma,
  title={Lingma swe-gpt: An open development-process-centric language model for automated software improvement},
  author={Ma, Yingwei and Cao, Rongyu and Cao, Yongchang and Zhang, Yue and Chen, Jue and Liu, Yibo and Liu, Yuchen and Li, Binhua and Huang, Fei and Li, Yongbin},
  journal={arXiv preprint arXiv:2411.00622},
  year={2024}
}

@article{chen2021evaluating,
  title={Evaluating large language models trained on code},
  author={Chen, Mark and Tworek, Jerry and Jun, Heewoo and Yuan, Qiming and Pinto, Henrique Ponde De Oliveira and Kaplan, Jared and Edwards, Harri and Burda, Yuri and Joseph, Nicholas and Brockman, Greg and others},
  journal={arXiv preprint arXiv:2107.03374},
  year={2021}
}

@article{austin2021program,
  title={Program synthesis with large language models},
  author={Austin, Jacob and Odena, Augustus and Nye, Maxwell and Bosma, Maarten and Michalewski, Henryk and Dohan, David and Jiang, Ellen and Cai, Carrie and Terry, Michael and Le, Quoc and others},
  journal={arXiv preprint arXiv:2108.07732},
  year={2021}
}

@article{zheng2023judging,
  title={Judging llm-as-a-judge with mt-bench and chatbot arena},
  author={Zheng, Lianmin and Chiang, Wei-Lin and Sheng, Ying and Zhuang, Siyuan and Wu, Zhanghao and Zhuang, Yonghao and Lin, Zi and Li, Zhuohan and Li, Dacheng and Xing, Eric and others},
  journal={Advances in Neural Information Processing Systems},
  volume={36},
  pages={46595--46623},
  year={2023}
}

@article{fu2023gptscore,
  title={Gptscore: Evaluate as you desire},
  author={Fu, Jinlan and Ng, See-Kiong and Jiang, Zhengbao and Liu, Pengfei},
  journal={arXiv preprint arXiv:2302.04166},
  year={2023}
}

@inproceedings{chen2024mllm,
  title={Mllm-as-a-judge: Assessing multimodal llm-as-a-judge with vision-language benchmark},
  author={Chen, Dongping and Chen, Ruoxi and Zhang, Shilin and Wang, Yaochen and Liu, Yinuo and Zhou, Huichi and Zhang, Qihui and Wan, Yao and Zhou, Pan and Sun, Lichao},
  booktitle={Forty-first International Conference on Machine Learning},
  year={2024}
}

@misc{zhuge2025agentasajudge,
title={Agent-as-a-Judge: Evaluating Agents with Agents},
author={Mingchen Zhuge and Changsheng Zhao and Dylan R. Ashley and Wenyi Wang and Dmitrii Khizbullin and Yunyang Xiong and Zechun Liu and Ernie Chang and Raghuraman Krishnamoorthi and Yuandong Tian and Yangyang Shi and Vikas Chandra and J{\"u}rgen Schmidhuber},
year={2025},
url={https://openreview.net/forum?id=DeVm3YUnpj}
}

@inproceedings{peng2024domain,
  title={Domain knowledge matters: Improving prompts with fix templates for repairing python type errors},
  author={Peng, Yun and Gao, Shuzheng and Gao, Cuiyun and Huo, Yintong and Lyu, Michael},
  booktitle={Proceedings of the 46th ieee/acm international conference on software engineering},
  pages={1--13},
  year={2024}
}

@inproceedings{chow2024pyty,
  title={Pyty: Repairing static type errors in python},
  author={Chow, Yiu Wai and Di Grazia, Luca and Pradel, Michael},
  booktitle={Proceedings of the IEEE/ACM 46th International Conference on Software Engineering},
  pages={1--13},
  year={2024}
}

@inproceedings{oh2024towards,
  title={Towards Effective Static Type-Error Detection for Python},
  author={Oh, Wonseok and Oh, Hakjoo},
  booktitle={Proceedings of the 39th IEEE/ACM International Conference on Automated Software Engineering},
  pages={1808--1820},
  year={2024}
}

@article{shi2024efficient,
  title={Efficient and green large language models for software engineering: Vision and the road ahead},
  author={Shi, Jieke and Yang, Zhou and Lo, David},
  journal={ACM Transactions on Software Engineering and Methodology},
  year={2024},
  publisher={ACM New York, NY}
}

@article{cruz2024innovating,
  title={Innovating for Tomorrow: The Convergence of Software Engineering and Green AI},
  author={Cruz, Lu{\'\i}s and Franch Gutierrez, Xavier and Mart{\'\i}nez-Fern{\'a}ndez, Silverio},
  journal={ACM Transactions on Software Engineering and Methodology},
  year={2024},
  publisher={ACM New York, NY}
}

@article{cohen2009pearson,
  title={Pearson correlation coefficient},
  author={Cohen, Israel and Huang, Yiteng and Chen, Jingdong and Benesty, Jacob and Benesty, Jacob and Chen, Jingdong and Huang, Yiteng and Cohen, Israel},
  journal={Noise reduction in speech processing},
  pages={1--4},
  year={2009},
  publisher={Springer}
}

@article{terragni2025future,
  title={The Future of AI-Driven Software Engineering},
  author={Terragni, Valerio and Vella, Annie and Roop, Partha and Blincoe, Kelly},
  journal={ACM Transactions on Software Engineering and Methodology},
  year={2025},
  publisher={ACM New York, NY}
}

@article{randell2025looking,
  title={Looking Back on Recovery Blocks and Conversations},
  author={Randell, Brian and Xu, Jie},
  journal={IEEE Transactions on Software Engineering},
  year={2025},
  publisher={IEEE}
}

@inproceedings{horton2019dockerizeme,
  title={Dockerizeme: Automatic inference of environment dependencies for python code snippets},
  author={Horton, Eric and Parnin, Chris},
  booktitle={2019 IEEE/ACM 41st International Conference on Software Engineering (ICSE)},
  pages={328--338},
  year={2019},
  organization={IEEE}
}

@inproceedings{jin2024pyanalyzer,
  title={PyAnalyzer: An Effective and Practical Approach for Dependency Extraction from Python Code},
  author={Jin, Wuxia and Xu, Shuo and Chen, Dawei and He, Jiajun and Zhong, Dinghong and Fan, Ming and Chen, Hongxu and Zhang, Huijia and Liu, Ting},
  booktitle={Proceedings of the IEEE/ACM 46th International Conference on Software Engineering},
  pages={1--12},
  year={2024}
}

@inproceedings{ye2022knowledge,
  title={Knowledge-based environment dependency inference for Python programs},
  author={Ye, Hongjie and Chen, Wei and Dou, Wensheng and Wu, Guoquan and Wei, Jun},
  booktitle={Proceedings of the 44th International Conference on Software Engineering},
  pages={1245--1256},
  year={2022}
}

@inproceedings{wang2020watchman,
  title={Watchman: Monitoring dependency conflicts for python library ecosystem},
  author={Wang, Ying and Wen, Ming and Liu, Yepang and Wang, Yibo and Li, Zhenming and Wang, Chao and Yu, Hai and Cheung, Shing-Chi and Xu, Chang and Zhu, Zhiliang},
  booktitle={Proceedings of the ACM/IEEE 42nd international conference on software engineering},
  pages={125--135},
  year={2020}
}

@inproceedings{zhong2024can,
  title={Can llm replace stack overflow? a study on robustness and reliability of large language model code generation},
  author={Zhong, Li and Wang, Zilong},
  booktitle={Proceedings of the AAAI Conference on Artificial Intelligence},
  volume={38},
  number={19},
  pages={21841--21849},
  year={2024}
}

@article{jin2024llms,
  title={From llms to llm-based agents for software engineering: A survey of current, challenges and future},
  author={Jin, Haolin and Huang, Linghan and Cai, Haipeng and Yan, Jun and Li, Bo and Chen, Huaming},
  journal={arXiv preprint arXiv:2408.02479},
  year={2024}
}

@inproceedings{siddiq2022securityeval,
  title={SecurityEval dataset: mining vulnerability examples to evaluate machine learning-based code generation techniques},
  author={Siddiq, Mohammed Latif and Santos, Joanna CS},
  booktitle={Proceedings of the 1st International Workshop on Mining Software Repositories Applications for Privacy and Security},
  pages={29--33},
  year={2022}
}

@article{fu2024constrained,
  title={Constrained decoding for secure code generation},
  author={Fu, Yanjun and Baker, Ethan and Ding, Yu and Chen, Yizheng},
  journal={arXiv preprint arXiv:2405.00218},
  year={2024}
}

@inproceedings{tony2023llmseceval,
  title={Llmseceval: A dataset of natural language prompts for security evaluations},
  author={Tony, Catherine and Mutas, Markus and Ferreyra, Nicol{\'a}s E D{\'\i}az and Scandariato, Riccardo},
  booktitle={2023 IEEE/ACM 20th International Conference on Mining Software Repositories (MSR)},
  pages={588--592},
  year={2023},
  organization={IEEE}
}

@inproceedings{mukherjee2021fixing,
  title={Fixing dependency errors for Python build reproducibility},
  author={Mukherjee, Suchita and Almanza, Abigail and Rubio-Gonz{\'a}lez, Cindy},
  booktitle={Proceedings of the 30th ACM SIGSOFT international symposium on software testing and analysis},
  pages={439--451},
  year={2021}
}

@inproceedings{wang2022smartpip,
  title={smartpip: A smart approach to resolving python dependency conflict issues},
  author={Wang, Chao and Wu, Rongxin and Song, Haohao and Shu, Jiwu and Li, Guoqing},
  booktitle={Proceedings of the 37th IEEE/ACM International Conference on Automated Software Engineering},
  pages={1--12},
  year={2022}
}

@article{gore2023syntax,
  title={Syntax Error Detection and Correction in Python Code using ML.},
  author={Gore, Deipali Vikram and Binoj, Mahima and Borate, Sayali and Devnani, Ritu and Gopale, Sakshi},
  journal={Grenze International Journal of Engineering \& Technology (GIJET)},
  volume={9},
  number={2},
  year={2023}
}

@inproceedings{santos2018syntax,
  title={Syntax and sensibility: Using language models to detect and correct syntax errors},
  author={Santos, Eddie Antonio and Campbell, Joshua Charles and Patel, Dhvani and Hindle, Abram and Amaral, Jos{\'e} Nelson},
  booktitle={2018 IEEE 25th International Conference on Software Analysis, Evolution and Reengineering (SANER)},
  pages={311--322},
  year={2018},
  organization={IEEE}
}

@article{vassallo2020developers,
  title={How developers engage with static analysis tools in different contexts},
  author={Vassallo, Carmine and Panichella, Sebastiano and Palomba, Fabio and Proksch, Sebastian and Gall, Harald C and Zaidman, Andy},
  journal={Empirical Software Engineering},
  volume={25},
  pages={1419--1457},
  year={2020},
  publisher={Springer}
}

@article{su2025learn,
  title={Learn-by-interact: A Data-Centric Framework for Self-Adaptive Agents in Realistic Environments},
  author={Su, Hongjin and Sun, Ruoxi and Yoon, Jinsung and Yin, Pengcheng and Yu, Tao and Ar{\i}k, Sercan {\"O}},
  journal={arXiv preprint arXiv:2501.10893},
  year={2025}
}

@inproceedings{zhang2024autocoderover,
author = {, Yuntong and Ruan, Haifeng and Fan, Zhiyu and Roychoudhury, Abhik},
title = {AutoCodeRover: Autonomous Program Improvement},
year = {2024},
isbn = {9798400706127},
publisher = {Association for Computing Machinery},
address = {New York, NY, USA},
url = {https://doi.org/10.1145/3650212.3680384},
doi = {10.1145/3650212.3680384},
booktitle = {Proceedings of the 33rd ACM SIGSOFT International Symposium on Software Testing and Analysis},
pages = {1592–1604},
numpages = {13},
location = {Vienna, Austria},
series = {ISSTA 2024}
}

@article{ruan2024specrover,
  title={Specrover: Code intent extraction via llms},
  author={Ruan, Haifeng and Zhang, Yuntong and Roychoudhury, Abhik},
  journal={arXiv preprint arXiv:2408.02232},
  year={2024}
}

@article{kong2025demystifying,
  title={Demystifying Memorization in LLM-Based Program Repair via a General Hypothesis Testing Framework},
  author={Kong, Jiaolong and Xie, Xiaofei and Liu, Shangqing},
  journal={Proceedings of the ACM on Software Engineering},
  volume={2},
  number={FSE},
  pages={2712--2734},
  year={2025},
  publisher={ACM New York, NY, USA}
}

@inproceedings{chen2024promise,
  title={Promise and Peril of Collaborative Code Generation Models: Balancing Effectiveness and Memorization},
  author={Chen, Zhi and Jiang, Lingxiao},
  booktitle={Proceedings of the 39th IEEE/ACM International Conference on Automated Software Engineering},
  pages={493--505},
  year={2024}
}

@inproceedings{yang2024unveiling,
  title={Unveiling memorization in code models},
  author={Yang, Zhou and Zhao, Zhipeng and Wang, Chenyu and Shi, Jieke and Kim, Dongsun and Han, Donggyun and Lo, David},
  booktitle={Proceedings of the IEEE/ACM 46th International Conference on Software Engineering},
  pages={1--13},
  year={2024}
}

@inproceedings{liu2024repobench,
title={RepoBench: Benchmarking Repository-Level Code Auto-Completion Systems},
author={Tianyang Liu and Canwen Xu and Julian McAuley},
booktitle={The Twelfth International Conference on Learning Representations},
year={2024},
url={https://openreview.net/forum?id=pPjZIOuQuF}
}

@inproceedings{li2024evocodebench,
title={EvoCodeBench: An Evolving Code Generation Benchmark with Domain-Specific Evaluations},
author={Jia Li and Ge Li and Xuanming Zhang and Yunfei Zhao and Yihong Dong and Zhi Jin and Binhua Li and Fei Huang and Yongbin Li},
booktitle={The Thirty-eight Conference on Neural Information Processing Systems Datasets and Benchmarks Track},
year={2024},
url={https://openreview.net/forum?id=kvjbFVHpny}
}
